\documentclass[aps,prd,onecolumn,eqsecnum,amsmath,nofootinbib,preprintnumbers,superscriptaddress]{revtex4-2}

\usepackage[dvipsnames]{xcolor}
\usepackage{graphicx,float}
\usepackage{amssymb,theorem,mathrsfs,times}
\usepackage{bm,multirow,mathtools,dsfont,setspace}
\usepackage[colorlinks,linkcolor=blue,anchorcolor=blue,citecolor=blue]{hyperref}
\usepackage{ulem,cancel}
\usepackage[justification=raggedright,singlelinecheck=true]{caption}
\usepackage{subcaption}
\usepackage{ragged2e}
\usepackage{revsymb}

\begin{document}
\title{Global structure and stability of Kerr-Bertotti-Robinson spacetime}
\preprint{\hfill {\small {ICTS-USTC/PCFT-26-60}}}
\date{\today}

\author{Yu-Sen Zhou}
\email{zhou\_ys@mail.ustc.edu.cn}

\affiliation{Interdisciplinary Center for Theoretical Study and Department of Modern Physics,\\
	University of Science and Technology of China, Hefei, Anhui 230026, China}

\author{Liang-Bi Wu}
\email{wulb@ucas.ac.cn}
\affiliation{School of Fundamental Physics and Mathematical Sciences, Hangzhou Institute for Advanced Study, UCAS, Hangzhou 310024, China}

\author{Ming-Fei Ji}
\email{jimingfei@mail.ustc.edu.cn}

\affiliation{Interdisciplinary Center for Theoretical Study and Department of Modern Physics,\\
	University of Science and Technology of China, Hefei, Anhui 230026, China}

\author{Wen-Tao Fu}
\email{fuwentao2024@mail.ustc.edu.cn}

\affiliation{Interdisciplinary Center for Theoretical Study and Department of Modern Physics,\\
	University of Science and Technology of China, Hefei, Anhui 230026, China}

\author{Li-Ming Cao}
\email{caolm@ustc.edu.cn}
\affiliation{Interdisciplinary Center for Theoretical Study and Department of Modern Physics,\\
	University of Science and Technology of China, Hefei, Anhui 230026, China}
\affiliation{Peng Huanwu Center for Fundamental Theory, Hefei, Anhui 230026, China}

\author{Rong-Gen Cai}
\email{caironggen@nbu.edu.cn}
\affiliation{Institute of Fundamental Physics and Quantum Technology, \& School of Physical Science and Technology, Ningbo University, Ningbo 315211, China}

\begin{abstract}
The surface $r=\infty$ of the Kerr-Bertotti-Robinson spacetime is not a genuine boundary. We construct its natural analytic extension and show that $r=+\infty$ of one KBR region is smoothly connected to $r=-\infty$ of a neighboring one. Repeated continuation produces an infinite chain of regions connected by wormhole-like bridges and exposes the neighboring ring singularity without an intervening horizon, which violates the weak cosmic censorship conjecture. We then study a test massless scalar field on a two-universe scattering segment to probe the stability of the spacetime. For axisymmetric perturbations, we analytically establish purely imaginary growing quasinormal modes for every $\ell$ and trace their origin to the chronology-violating region. In the $(\ell,m)=(2,2)$ sector, unstable branches occur for sufficiently large rotation and sufficiently small magnetic field. Their marginal real modes obey an exact horizon-flux balance, supporting a black-hole-bomb interpretation in which superradiant extraction is amplified by trapping within the double-barrier potential. The same cavity also supports families of weakly damped modes and may produce echo-like responses.

\end{abstract}

\maketitle

\section{Introduction}\label{sec-introduction}

Astrophysical black holes are often immersed in strong electromagnetic fields. Two classic Einstein-Maxwell backgrounds provide exact settings in which such fields are incorporated directly into the geometry. The Bonnor-Melvin universe describes a self-gravitating magnetic field~\cite{Bonnor:1954tis,Melvin:1963qx} and underlies the Ernst-Wild family of magnetized Kerr-Newman solutions~\cite{Ernst:1976mzr,Ernst:1976bsr}, whereas the Bertotti-Robinson (BR) universe describes a homogeneous electromagnetic environment~\cite{Bertotti:1959pf,Robinson:1959ev}. A rotating black-hole-like geometry in the latter setting is provided by the Kerr-Bertotti-Robinson (KBR) solution~\cite{Podolsky:2025tle}, an exact Petrov type-D solution of the Einstein-Maxwell equations. Broader charged and accelerating families and related electromagnetic backgrounds have subsequently been constructed~\cite{Ovcharenko:2025cpm,Astorino:2025lih}, and the original KBR solution was recently identified as a special charged member of a more general Kerr-Newman-Bertotti-Robinson family~\cite{Ovcharenko:2026tos}. These developments have stimulated extensive studies of the phenomenology and mathematical structure of KBR and related spacetimes~\cite{Zeng:2025olq,Wang:2025vsx,Zeng:2025tji,Wang:2025bjf,Ali:2025beh,Zhang:2025ole,Liu:2025wwq,Li:2025rtf,Mirkhaydarov:2026fyn,Wang:2026czl,Mustafa:2026gly,Ahmed:2025ril,Siahaan:2025ngu,Gray:2025lwy,Barrientos:2026kdl,Siahaan:2026tuf,Lu:2026kcm,Hu:2026slp,Rehman:2026rzq,Wan:2026lca,Roy:2026poj,Hassanabadi:2026fzz,Huang:2026qzj,Ahmed:2026ozq,Singh:2026rbz}.

Despite this progress, the global interpretation of the KBR geometry remains subtle. In the Boyer-Lindquist-type coordinates, the surface $r=\infty$ is known not to represent null infinity~\cite{Podolsky:2025tle}. Null geodesics reach it within finite affine parameter, timelike observers can reach it within finite proper time, and the Weyl scalar $\Psi_2$ remains nonvanishing there. The surface therefore does not represent an asymptotic boundary, and a single KBR coordinate region does not describe the complete spacetime. This also obstructs the usual global notion of a black hole defined relative to an asymptotic boundary, although the outer Killing horizon can still be interpreted quasi-locally, for example within the isolated-horizon framework~\cite{Ashtekar:1998sp,ashtekar:2004cn}.

The natural continuation across this surface reveals a substantially richer global structure. The surface $r=+\infty$ of one KBR region is smoothly connected to $r=-\infty$ of a neighboring region, and repeated continuation produces an infinite chain of regions connected by wormhole-like bridges. The extension exposes the ring singularity of the neighboring region without an intervening horizon and also gives access to its chronology-violating negative-radius sector. The exposed singularity appears to challenge weak cosmic censorship conjecture, although the standard modern formulation concerns evolution from generic initial data and assumes an appropriate complete future null infinity~\cite{Wald:1997wa}, which is absent here. The global structure also depends nontrivially on the parameters. In particular, the gluing surface is timelike, null, or spacelike according to whether $A$ is positive, zero, or negative, respectively, and the extremal, static, and massless limits exhibit distinct causal structures.

These features make the perturbative properties of the extended geometry especially important. Quasinormal modes (QNMs) provide a natural probe of its response~\cite{Kokkotas:1999bd,Nollert:1999ji,Berti:2009kk,Konoplya:2011qq,Berti:2025hly}. We consider a test massless scalar field on a two-universe scattering segment extending from the outer horizon of one region, across the wormhole bridge, to the inner horizon of the neighboring region. In the axisymmetric sector, a self-adjoint formulation adapted from the Kerr analysis of Ref.~\cite{Dotti:2011ix} proves the existence of purely imaginary growing QNMs for every $\ell$. Their radial trapping region coincides with the equatorial chronology-violating region, strongly linking these modes to the causal pathology of the negative-radius sector. Since the scattering region is not globally hyperbolic, however, these growing modes do not establish a conventional dynamical instability generated from generic freely specified Cauchy data.

A qualitatively different phenomenon appears in the nonaxisymmetric $(\ell,m)=(2,2)$ sector. Unstable QNM branches occur for sufficiently large $a/M$ and sufficiently small $BM$, demonstrating that the growing spectrum is not restricted to the special axisymmetric construction. At the onset of instability, these branches cross the real axis at marginal modes for which the conserved radial Wronskian, together with the QNM boundary conditions, gives an exact balance between the flux terms at the two horizons. With the corresponding future-horizon orientations, the balance describes absorption at the outer horizon compensated by superradiant extraction from the neighboring inner horizon. The unstable branches emerge continuously from these marginal modes, while the effective potential forms an intermediate double-barrier cavity that supplies the feedback required for repeated amplification. These properties provide strong evidence for a black-hole-bomb-type mechanism~\cite{Press:1972zz,Cardoso:2004nk}. The cavity exists in a broader parameter region than the instability itself and also supports families of weakly damped modes, suggesting echo-like responses associated with the wormhole structure.

A concise account of the central results was presented in our short report~\cite{Zhou:2026tkm}. The present work provides the detailed geometric construction, the classification of the causal structures throughout parameter space and in the relevant limiting cases, and the analytical and numerical developments underlying the spectral results. Following the appearance of our short report~\cite{Zhou:2026tkm}, and while the present detailed study was being completed, the broader KNBR family was identified~\cite{Ovcharenko:2026tos}. The extension method developed here carries over naturally to the generic KNBR family, as also demonstrated in parallel in Ref.~\cite{Ovcharenko:2026ooh}, although a special parameter sector requires additional treatment and is currently under investigation.

The paper is organized as follows. Section~\ref{sec-geometry} introduces the KBR geometry and constructs its natural extension. Section~\ref{sec-causal} classifies the resulting causal structures and discusses the relevant special limits. Section~\ref{sec-waves} formulates the scalar scattering problem, develops the effective-potential picture, establishes the axisymmetric instability, and presents the Heun construction. Section~\ref{sec-spectra} presents the numerical spectra and their physical interpretation. Section~\ref{sec-conclusion} summarizes the results and discusses open questions.

\section{The Natural Extension}\label{sec-geometry}
The line element of the Kerr-Bertotti-Robinson (KBR) solution~\cite{Podolsky:2025tle} is
\begin{eqnarray}\label{ds}
	\mathrm{d}s^{2}=\frac{1}{\Omega^{2}}\Big[-\frac{Q}{\rho^{2}}\Big(\mathrm{d}t-a\Delta_xC\mathrm{d}\varphi\Big)^{2}+\frac{\rho^{2}}{Q}\mathrm{d}r^{2}+\frac{\rho^{2}}{P\Delta_x}\mathrm{d}x^{2}+\frac{P\Delta_x}{\rho^{2}}\Big(a\mathrm{d}t-\rho_0^2C\mathrm{d}\varphi\Big)^{2}
	\Big]\, ,
\end{eqnarray}
where the metric functions are given as follows
\begin{eqnarray}
	\rho_0^2(r)&=& r^{2}+a^{2}\, ,\nonumber\\
	\rho^{2}(r,x) &=& r^{2}+a^{2}x^{2}\, ,\nonumber\\
	P(x) &=& 1+\Big(\frac{1}{C}-1\Big)x^{2}\, ,\nonumber\\
	Q(r) &=& (1+B^{2}r^{2})\Delta\, ,\nonumber\\
	\Omega^{2}(r,x) &=& (1+B^{2}r^{2})-B^{2}\Delta x^{2}\, ,\nonumber\\
	\Delta(r) &=& A r^{2}-2\mu r+a^{2}\, ,\nonumber\\
	\Delta_x(x)&=&1-x^2\, ,
\end{eqnarray}
and
\begin{eqnarray}
\label{metricfunctions2}
	I_{1}&=&1-\frac{1}{2}B^{2}a^{2}\, ,\quad
	I_{2}=1-B^{2}a^{2}\, ,\quad
	\mu=M\frac{I_2}{I_1}\, ,\nonumber\\
	A&=&1-B^{2}M^{2}\frac{I_{2}}{I_{1}^{2}}\, ,\quad
	C=\Big[1+B^{2}\Big(\frac{M^{2}I_{2}}{I_{1}^{2}}-a^{2}\Big)\Big]^{-1}\, .
\end{eqnarray}
Here $M$, $a$, and $B$ denote the mass parameter, rotation parameter, and strength of the external uniform magnetic field, respectively. We use $x=\cos\theta$ and explicitly include the conicity factor $C$ to remove conical defects on the symmetry axis, so that $\varphi\in[0,2\pi)$~\cite{Destounis:2020pjk, Xiong:2023usm, Chen:2024rov}. The complex electromagnetic potential is given by
\begin{eqnarray}
	A_\mu \mathrm{d}x^\mu=\frac{\mathrm{e}^{\mathrm{i}\nu}}{2B}\Bigg[\Omega_{,r}\frac{a\mathrm{d}t-\rho_0^2\mathrm{d}\varphi}{r+\mathrm{i}ax}-\mathrm{i}\Omega_{,x}\frac{\mathrm{d}t-a\Delta_x\mathrm{d}\varphi}{r+\mathrm{i} a x}+(\Omega-1)\mathrm{d}\varphi
		\Bigg]\, .
\end{eqnarray}
Here $\nu$ is the duality-rotation parameter, and the physical gauge potential is obtained as $A_\mu^{\text{real}}=2\mathrm{Re}(A_\mu)$. The corresponding field strength $F=\mathrm{d}A^{\text{real}}$, together with the metric~(\ref{ds}), satisfies the Einstein-Maxwell equations~\cite{Podolsky:2025tle}.

\begin{figure}[htbp]
	\centering
	\includegraphics[width=0.48\textwidth]{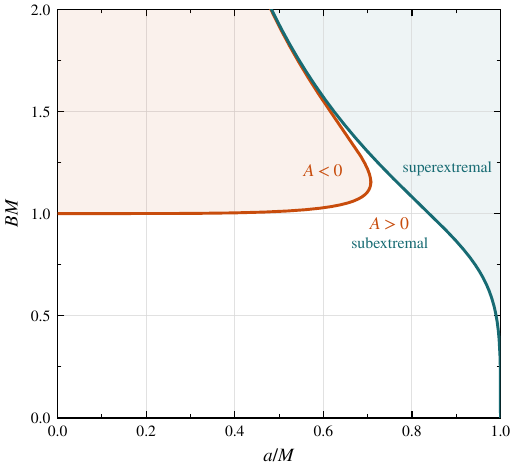}
	\caption{\justifying Parameter space in the $(a/M,BM)$ plane. The teal curve denotes the extremal bound, with the subextremal region on its smaller-$a$ side. The orange curve marks the critical locus $A=0$; the region to its right, i.e. the larger-$a$ side, has $A>0$. As $a\to0$, the lower orange branch approaches $BM=1$, while the upper orange branch and the extremal curve both scale as $B\sim a^{-1}$.}
	\label{fig:regime}
\end{figure}

The global structure depends on the sign of $A$ in eqs.(\ref{metricfunctions2}) and on extremality. The corresponding parameter space is shown in Fig. \ref{fig:regime}. The extremal condition is given by 
\begin{equation}
\label{conditionextremal}
\mu^2=Aa^2\, ,
\end{equation}
where two real roots of $Q$ coincide. As $a/M\to0$, the lower branch of the $A=0$ curve approaches $BM=1$ as
\[
BM=1+\frac{1}{8}\left(\frac{a}{M}\right)^4
+\mathcal{O}\left[\left(\frac{a}{M}\right)^6\right],
\]
while its upper branch and the extremal curve both diverge as
\[
BM=\frac{M}{a}-\frac{1}{8}\frac{a}{M}
+\mathcal{O}\left[\left(\frac{a}{M}\right)^3\right].
\]
They first differ at cubic order, with coefficients $-13/128$ and $-9/128$, respectively. The two branches of the $A=0$ curve meet at
$a/M=1/\sqrt{2}$ and $BM=2/\sqrt{3}$, whereas the extremal curve approaches
the Kerr extremal point $(a/M,BM)=(1,0)$ as
\[
BM=
\sqrt{2}\left[1-\left(\frac{a}{M}\right)^2\right]^{1/4}
+\mathcal{O}\left(
\left[1-\left(\frac{a}{M}\right)^2\right]^{3/4}
\right).
\]

\subsection{Generic subextremal $A>0$ extension}
The real roots of $Q(r)$ locate the Killing horizons. In the subextremal $A>0$ case shown in Fig. \ref{fig:regime}, there are two positive roots, $r_{\mathrm{o}}$ and $r_{\mathrm{i}}$, corresponding to the outer and inner horizons. The remaining roots of $Q(r)$ are $r_\pm=\pm\mathrm{i}/B$. We first construct the extension in the generic subextremal sector with $A>0$ and $r_{\mathrm{i}}<r_{\mathrm{o}}$. The remaining causal sectors and special limits are discussed in Sec.~\ref{sec-causal}. 

The Weyl tensor of the KBR spacetime is of Petrov type D and admits two repeated principal null directions. The outgoing one is represented by
\begin{equation}
\ell^\mu\partial_\mu=\Omega^2\left(\frac{\rho_0^2}{Q}\partial_t+\partial_r+\frac{a}{QC}\partial_\varphi\right)\,,
\end{equation}
which, under this normalization, generates an affinely parametrized null geodesic congruence of the physical metric~\eqref{ds}. Consider an integral curve of this congruence, parameterized by an affine parameter $s$. The geodesic satisfies
\begin{eqnarray}
    \dot t-\dot r_*=\dot x=\dot{\varphi}-\dot{\psi}=0\, ,
\end{eqnarray}
where a dot denotes $\mathrm{d}/\mathrm{d}s$ and the tortoise coordinate $r_*$ and the dragged azimuthal coordinate $\psi$ are defined as
\begin{eqnarray}
	r_*&=&\int\frac{\rho_0^2}{Q}\mathrm{d}r\nonumber\\
    &=&\frac{1}{2\kappa_{\text{o}}}\left[\ln\left|\frac{B\left(r-r_{\text{o}}\right)}{\sqrt{1+B^2r^2}}\right|+Br_{\text{o}}\arctan\left(\frac{1}{Br}\right)\right]+\frac{1}{2\kappa_{\text{i}}}\left[\ln\left|\frac{B\left(r-r_{\text{i}}\right)}{\sqrt{1+B^2r^2}}\right|+Br_{\text{i}}\arctan\left(\frac{1}{Br}\right)\right]\nonumber\\&+&\frac{1}{AB}\Big[\arctan(Br)-\frac{\pi}{2}\Big]\, ,\nonumber\\
	\psi&=&\int\frac{a}{\rho_0^2C}\mathrm{d}r_*=\int\frac{a}{QC}\mathrm{d}r\nonumber\\
	&=&\frac{\varpi_{\text{o}}}{2\kappa_{\text{o}}}\left[\ln\left|\frac{B(r-r_{\text{o}})}{\sqrt{1+B^2r^2}}\right|+Br_{\text{o}}\arctan\left(\frac{1}{Br}\right)\right]+\frac{\varpi_{\text{i}}}{2\kappa_{\text{i}}}\left[\ln\left|\frac{B(r-r_{\text{i}})}{\sqrt{1+B^2r^2}}\right|+Br_{\text{i}}\arctan\Big(\frac{1}{Br}\Big)\right]\, ,
\end{eqnarray}
in which
\begin{eqnarray}\label{kappaDef}
	\kappa_j=\frac{Q'}{2\rho_0^2}\Bigg|_{r=r_j}\, , \quad j=\text{o},\text{i},+,-\, .
\end{eqnarray}
Here $\kappa_{\mathrm{o}}$ and $\kappa_{\mathrm{i}}$ are the surface gravities of the two real Killing horizons, while $\kappa_{\pm}$ are defined analogously at the complex roots for notational convenience. The additive constant in $r_*$ is chosen such that it approaches zero at $r\to+\infty$. The affine parameter along the ray satisfies \begin{equation}
\frac{\mathrm{d}s}{\mathrm{d}r}=\frac{1}{E\Omega^2}=\mathcal{O}(r^{-2})\,,
\end{equation}
where $E$ denotes the conserved energy of the light ray. The full integrated expression for the affine parameter, after choosing the integration constant such that $s(r=+\infty)=0$, is
	\begin{eqnarray}
		s &=&
		\begin{dcases}
			-\frac{1}{E B \sqrt{\Xi(x)}}\arctan\left(\frac{\sqrt{\Xi(x)}}{B\left[(1-Ax^2)r+\mu x^2\right]}\right)\,, & |x|<1\,; \\
			-\frac{1}{E B^2 \left[(1-A)r+\mu\right]}\,,                                                              & |x|=1\,,
		\end{dcases}
	\end{eqnarray}
	where
	\begin{eqnarray}
		\Xi(x)=(1-x^2)\left[1+B^2(\mu^2-Aa^2)x^2\right]\,.\nonumber
	\end{eqnarray}
The facts that light can reach $r=\infty$ within finite affine parameter and that $\Psi_2$ approaches a nonzero constant~\cite{Podolsky:2025tle} as $r\to\infty$ both indicate that this surface is not a genuine boundary.

To investigate the causal structure near $r=\infty$, it is convenient to introduce a compactified radial coordinate $y$. To guarantee that the resulting metric component $g_{yy}$ is regular at $r\to\infty$, $y$ should satisfy $\mathrm{d}y/\mathrm{d}r\sim r^{-2}+\mathcal{O}(r^{-3})$; hence any regular radial defining function is proportional to $1/r$ at leading order, up to a smooth reparametrization. Specifically, we choose $y=-r_{\text{o}}/r$. This coordinate transformation maps
\begin{equation}
r=\infty,\quad r_{\text{o}},\quad r_{\text{i}},\quad r_{+},\quad r_{-}
\quad\longmapsto\quad y=
0,\quad -1,\quad -\frac{r_{\text{o}}}{r_{\text{i}}},\quad \mathrm{i}Br_{\text{o}},\quad -\mathrm{i}Br_{\text{o}}\,,
\end{equation}
respectively. In terms of $(t,y,x,\varphi)$, the line element~\eqref{ds} becomes
\begin{eqnarray}
	\mathrm{d}s^2&=&\frac{1}{\widehat{\Omega}^2}\Bigg[-\frac{\widehat{Q}}{\widehat{\rho}^2}\Big(\mathrm{d}t-a\Delta_xC\mathrm{d}\varphi\Big)^2+\frac{\widehat{\rho}^2}{\widehat{Q}}r_{\text{o}}^2\mathrm{d}y^2+\frac{\widehat{\rho}^2}{P\Delta_x}\mathrm{d}x^2+\frac{P\Delta_x}{\widehat{\rho}^2}\Big(ay^2\mathrm{d}t-\widehat{\rho}_0^2C\mathrm{d}\varphi\Big)^2\Bigg]\, ,
\end{eqnarray}
where
\begin{eqnarray}
	\widehat{\rho}_0^2&=&y^2\rho_0^2(-r_{\text{o}}/y)=r_{\text{o}}^2+a^2y^2\,,\nonumber\\
	\widehat{\rho}^2&=&y^2\rho^2(-r_{\text{o}}/y,x)=r_{\text{o}}^2+a^2x^2y^2\,,\nonumber\\
	\widehat{Q}&=&y^4Q(-r_{\text{o}}/y)=(y^2+B^2r_{\text{o}}^2)\widehat{\Delta}\,,\nonumber\\
	\widehat{\Delta}&=&y^2\Delta(-r_{\text{o}}/y)=Ar_{\text{o}}^2+2\mu r_{\text{o}}y+a^2y^2\,,\nonumber\\
	\widehat{\Omega}&=&|y|\Omega(-r_{\text{o}}/y,x)=\sqrt{(y^2+B^2r_{\text{o}}^2)-B^2\widehat{\Delta}x^2}\,.\nonumber
\end{eqnarray}
These hatted quantities are smooth at $y=0$, and hence the metric extends smoothly across this surface. We choose $\Omega$ to have the same sign in the two KBR regions. Smooth continuation of the electromagnetic field then requires a shift of the duality angle by $\pi$, $\nu\to\nu+\pi$, in the neighboring region. Equivalently, one may instead keep $\nu$ fixed and let $\Omega$ change sign across $y=0$. The geometry is unchanged because the metric depends only on $\Omega^2$. The hypersurface $y=0$ is timelike, and the induced metric and extrinsic curvature have the same limits on the two sides, so no matter shell is introduced~\cite{Israel:1966rt,Poisson:2009pwt,Mars:1993mj}. Since the metric and Maxwell field are analytic in $y$ at $y=0$, their continuation is locally unique up to regular coordinate reparametrizations and electromagnetic gauge transformations~\cite{Choquet-Bruhat:1969ywq,Friedrich:2000qv,Rendall:2000pk,Friedrich:1998xt,Friedrich:2009tq}. In this sense the extension is natural rather than an arbitrary gluing.

We may therefore continue the spacetime across $y=0$. The corresponding tortoise coordinate $\bar r_*$ is normalized so that $\bar r_*\to0$ as $\bar r\to-\infty$, matching $r_*\to0$ at $r\to+\infty$. In the extended region we denote the radial coordinate by $\bar r=-r_{\mathrm o}/y$. The surface $r=+\infty$ of the first region is smoothly connected to $\bar r=-\infty$ of a neighboring KBR region with the same metric parameters. Since the smooth Maxwell continuation shifts $\nu$ by $\pi$, adjacent regions cannot be identified while keeping the electromagnetic field single valued. A period-two identification is possible after two continuations, but we keep the regions distinct.

The resulting geometry has a wormhole-like structure, with $y=0$ connecting the exterior of one KBR region to the negative-radius interior of its neighbor. Following the same null ray beyond $y=0$, one finds
\begin{equation}
\frac{\mathrm d s}{\mathrm d y}=\frac{r_{\mathrm o}}{E\widehat\Omega^2}=\mathcal O(y^{-2})
\end{equation}
as $y\to+\infty$. 	The affine parameter in the $y$ coordinate with $s(y=+\infty)=0$ is
	\begin{eqnarray}
		s &=&
		\begin{dcases}
			-\frac{1}{EB\sqrt{\Xi(x)}}
			\arctan\left(
			\frac{Br_{\text{o}}\sqrt{\Xi(x)}}{(1-a^2B^2x^2)y-\mu B^2r_{\text{o}}x^2}
			\right)\,,                                                  & |x|<1\,; \\
			-\frac{r_\text{o}}{E\left[(1-a^2B^2)y-\mu B^2r_\text{o}\right]}\,, & |x|=1\,.
		\end{dcases}
	\end{eqnarray}
Provided that the ray avoids the ring singularity and the isolated axial points $\{r=-\mu/(1-A),x=\pm1\}$ where $\Omega=0$, it can pass through $\bar r=0$ and reach $\bar r=+\infty$ within finite affine parameter. The same construction can then be iterated, producing an infinite chain of KBR regions.

As in Kerr, the ring singularity is located at $r=0$, $x=0$. Its leading curvature behavior can be written as
\begin{eqnarray}
    R_{abcd}R^{abcd}=\frac{48\mu^2\cos(6\alpha)}{\rho^6}+\mathcal O(\rho^{-5})\, ,\qquad \alpha=\arctan\frac{ax}{r}\, .
\end{eqnarray}
The neighboring region therefore contains a ring singularity with no horizon separating it from the original exterior. This challenges weak cosmic censorship conjecture. The standard modern formulation, however, concerns evolution from generic initial data and assumes an appropriate complete future null infinity~\cite{Wald:1997wa}, which is absent in the extended KBR geometry. The exposed singularity should therefore be interpreted together with the possibility that this exact geometry, if realizable through evolution at all, requires highly special initial data.

\subsection{Isometric embedding of the bridge}
\begin{figure}
	\includegraphics[width=0.7\columnwidth]{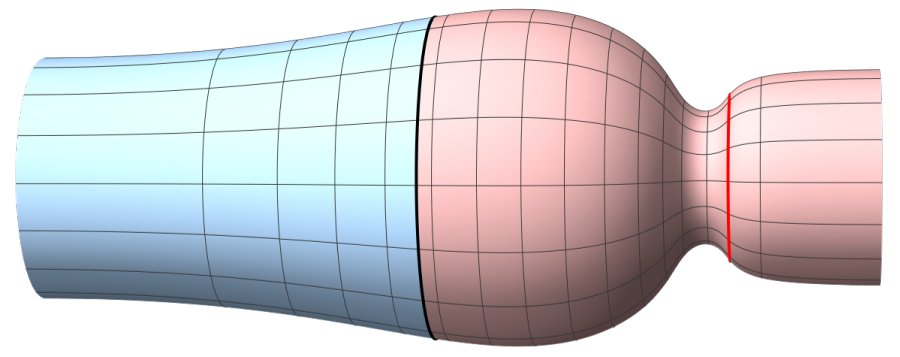}
	\caption{\justifying Isometric embedding of the fixed-$t$, fixed-$x=x_{0}=0.7$ spatial slice for $M=1$, $a=0.9$, and $B=0.6$. The gluing surface $r=+\infty\sim\bar r=-\infty$ and the surface $\bar r=0$ are indicated by the black and red circles, respectively. We take $x_{0}\neq0$ to avoid the ring singularity, and the parameters are chosen such that the slice avoids both the region containing closed timelike azimuthal curves and the region that does not admit an axisymmetric isometric embedding in Euclidean three-space. In the latter region, the circumferential radius varies too rapidly with proper radial distance to be realized by a Euclidean surface of revolution. An analogous ring singularity, chronology-violating region, and Euclidean-embedding obstruction also arise in the negative-$r$ extension of the Kerr spacetime~\cite{Boyer:1966qh, Carter:1968rr,Bardeen:1972fi}.}
	\label{fig:embed}
\end{figure}

The extension possesses a wormhole-like character, which can be visualized intrinsically on a fixed-$t$, fixed-$x$ spatial slice, as shown in Fig.~\ref{fig:embed}. 

The embedding is obtained as follows. We consider the fixed-$t$, fixed-$x=x_{0}\neq0$ two-dimensional slice. We choose $x_{0}\neq0$ to avoid the ring singularity. Since neither $r$ nor $\bar r$ covers the entire region across the gluing surface $r=+\infty\sim\bar r=-\infty$, we introduce a single compact coordinate $\chi$ by
\begin{equation}
    \begin{cases}
        r=r_{\mathrm{o}}\tan\chi\,, & \displaystyle\frac{\pi}{4}\leq\chi<\frac{\pi}{2}\,,\\
        \bar r=r_{\mathrm{o}}\tan\chi, & \displaystyle\frac{\pi}{2}<\chi\leq\pi+\arctan\left(\frac{r_{\mathrm{i}}}{r_{\mathrm{o}}}\right)\, .
    \end{cases}
\end{equation}
The value $\chi=\pi/2$ represents the gluing surface $r=+\infty\sim\bar r=-\infty$. The induced metric then takes the form
\begin{equation}
\mathrm{d}l^{2}=E(\chi)\mathrm{d}\chi^{2}+F(\chi)\mathrm{d}\varphi^{2}\, .\label{m1}
\end{equation}
We embed this intrinsic geometry isometrically into flat Euclidean three-space as a surface of revolution,
\begin{equation}
    \boldsymbol{X}(\chi\,,\varphi)=\left(\sqrt{F(\chi)}\cos\varphi\,,\sqrt{F(\chi)}\sin\varphi\,,Z(\chi)\right)\, .
\end{equation}
The metric induced on this surface by the flat Euclidean metric is
\begin{equation}
    \mathrm{d}l^2=\left[Z'(\chi)^{2}+\frac{F'(\chi)^{2}}{4F(\chi)}\right]\mathrm{d}\chi^{2}+F(\chi)\mathrm{d}\varphi^{2}\, .\label{m2}
\end{equation}
Comparing \eqref{m1} and \eqref{m2}, we obtain
\begin{equation}
    Z'(\chi)=\pm\sqrt{E(\chi)-\frac{F'(\chi)^{2}}{4F(\chi)}}\, .
\end{equation}
The embedding is therefore obtained by integrating
\begin{equation}
    Z(\chi)=Z(\chi_{0})\pm\int_{\chi_{0}}^{\chi}\sqrt{E(\widetilde{\chi})-\frac{F'(\widetilde{\chi})^{2}}{4F(\widetilde{\chi})}}\mathrm{d}\widetilde{\chi}\, .
\end{equation}
The construction is valid only where
\begin{equation}
    F(\chi)=g_{\varphi\varphi}>0\,,\quad E(\chi) \geq \frac{F'(\chi)^{2}}{4F(\chi)}\, .
\end{equation}
The first condition excludes the closed timelike azimuthal orbits. The second condition ensures that $Z'(\chi)$ is real and hence that the surface admits an axisymmetric isometric embedding in Euclidean three-space. Indeed, we introduce the proper radial distance $\mathfrak{s}$ and the circumferential radius $\mathfrak{R}$ by
\begin{equation}
    \mathrm{d}\mathfrak{s}=\sqrt{E(\chi)}\mathrm{d}\chi\,,
    \quad \mathfrak{R}(\chi)=\sqrt{F(\chi)}\, ,
\end{equation}
so that the circumference of an azimuthal circle is $2\pi \mathfrak{R}(\chi)$. One then finds
\begin{equation}
    1-\left(\frac{\mathrm{d}\mathfrak{R}}{\mathrm{d}\mathfrak{s}}\right)^{2}=\frac{1}{E(\chi)}\left[E(\chi)-\frac{F'(\chi)^{2}}{4F(\chi)}\right]=\left(\frac{\mathrm{d}Z}{\mathrm{d}\mathfrak{s}}\right)^{2}\, .
    \label{con}
\end{equation}
Therefore, the second condition can be alternatively understood as requiring the circumferential radius not to vary faster than the proper radial distance. It can be proved that these two conditions can always be satisfied by taking $|x_{0}|$ sufficiently close to, but strictly less than, unity.

\section{Global Causal Structures}\label{sec-causal}
The extension is not restricted to the generic subextremal $A>0$ sector shown in Fig.~\ref{fig:regime}. The sign of $A$ controls the causal character of the $r=\infty$ hypersurface, while extremality and the limits $a=0$ and $M=0$ introduce additional degeneracies. We now treat these cases explicitly. The case $B=0$ reduces to Kerr, for which $r\to+\infty$ is a genuine boundary and the present continuation is absent.

\subsection{Generic subextremal $A>0$ sector}
\begin{figure}[htbp]
	\centering
	\resizebox{0.5\columnwidth}{!}{
	\includegraphics{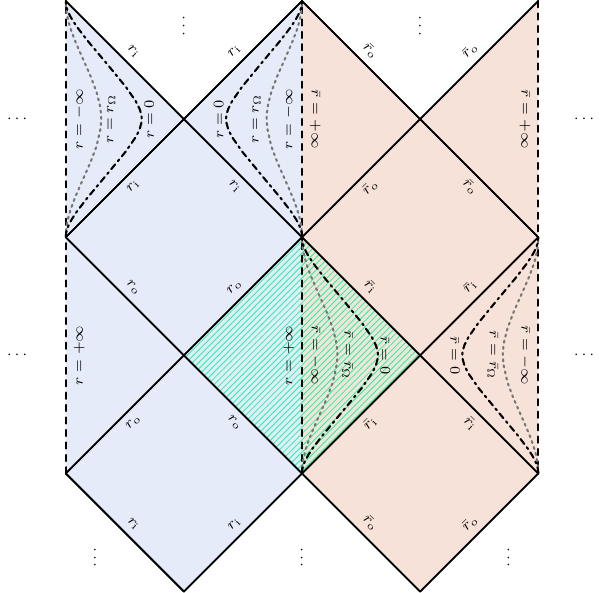}}
	\caption{\justifying Penrose diagram of the extended spacetime considered in this work. Only a local two-universe segment is displayed here, as the full diagram is obtained by infinitely repeating this minimal building block. The left and right colored regions correspond to the first and second universes, respectively, while the ellipses indicate the infinite continuation of the diagram by further gluing. The straight dashed lines denote the timelike surfaces $r=\pm\infty$ and $\bar r=\pm\infty$, across which neighboring universes are connected. The dash-dotted curves mark $r=0$ and $\bar r=0$, with ring singularities at $x=0$. The gray dotted curves indicate $r_\Omega=\bar r_\Omega=-\mu/(1-A)$, where $\Omega$ vanishes at $x=\pm1$. The corresponding Penrose diagrams on the axis and equatorial plane are obtained simply by truncating the extension at $r=r_\Omega$ and $r=0$, respectively, with $r=r_\Omega$ replaced by null infinity and $r=0$ by the singularity. The hatched diamond denotes the scattering region considered here.}
	\label{fig:kbr-penrose}
\end{figure}

For $A>0$ and $\mu^2>Aa^2$, the two real roots of $Q$ are positive and define the outer and inner Killing horizons. The surface $r=\infty$ is timelike and the continuation of Sec.~\ref{sec-geometry} glues it to $\bar r=-\infty$ of the neighboring region. Repeated continuation produces the timelike chain shown in Fig.~\ref{fig:kbr-penrose}.

\subsection{Extremal sector}
\begin{figure}[htbp]
	\centering
	\includegraphics{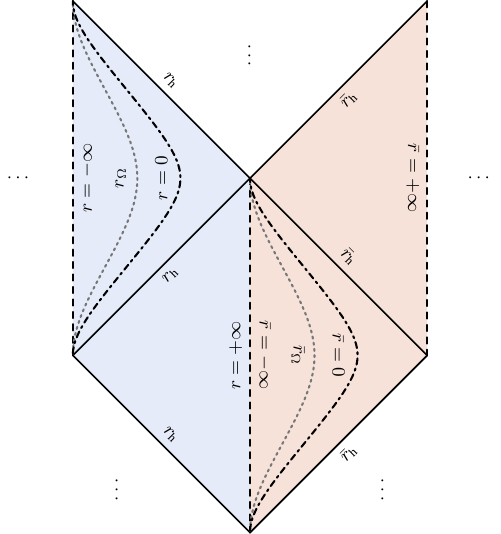}
	\caption{\justifying Schematic of the minimal building block of the full Penrose diagram for an extremal KBR geometry. The blue and red regions denote the first universe and the glued second universe, respectively. The ellipses indicate infinite continuation through further gluing, so that the full Penrose diagram is obtained by repeated copies of this basic building block. The vertical dashed lines represent the timelike gluing surface $r=+\infty\sim\bar r=-\infty$, while the dashed curves mark the ring singularities at $r=0$ and $\bar r=0$.}
	\label{fig:ex-penrose}
\end{figure}

% Representative exact-curve parameters used below: M=1, a=0.6,
% B=B_ext(0.6)=1.571348402637, satisfying mu^2=A a^2.
The extremal KBR geometry arises when the two real roots of $Q$ coalesce. This happens when the condition (\ref{conditionextremal}) is satisfied.
%\begin{eqnarray}
%	\mu^2=Aa^2\,.
%\end{eqnarray}
In this case
\begin{eqnarray}
	r_{\text{h}}=\frac{\mu}{A}\,,
	\quad
	Q(r)=A(1+B^2r^2)(r-r_{\text{h}})^2\,.
\end{eqnarray}
Thus the outer horizon $r=r_{\text{o}}$ and the inner horizon $r=r_{\text{i}}$ of the nonextremal spacetime merge into a single degenerate Killing horizon at $r=r_{\text{h}}$. The axial degeneration of the KBR coordinates remains at
\begin{equation}
	r_\Omega=-\frac{\mu}{1-A}\,,\label{rOm}
\end{equation}
where $\Omega=0$ at $x=\pm1$. The radial outgoing null rays can still reach $r=+\infty$ in finite affine parameter, as
\begin{eqnarray}
	\frac{\mathrm{d}s}{\mathrm{d}r}
	=\frac{1}{E\Omega^2}
	=\frac{1}{E(1-Ax^2)B^2r^2}+\mathcal{O}\left(\frac{1}{r^3}\right)\,.
\end{eqnarray}
Therefore $r=+\infty$ is again not a real boundary. We introduce the coordinate
\begin{eqnarray}
	y=-\frac{r_{\text{h}}}{r}\,.
\end{eqnarray}
The metric functions obtained from this coordinate transformation remain smooth and nonvanishing at $y=0$, so the extremal spacetime can again be glued to a second copy at $r=+\infty$, identified with $\bar r=-\infty$. Iterating this construction produces an infinite chain of extremal KBR exteriors joined by bridges. The resulting Penrose diagram is shown in Fig.~\ref{fig:ex-penrose}.

\subsection{Critical sector with $A=0$}
% Representative exact-curve parameters: M=1, a=0.6,
% B=b_-(0.6)=1.028920325349 (the smaller A=0 branch).
For $A=0$, the function $\Delta(r)$ degenerates from a quadratic
polynomial to a linear one, and we have
\begin{eqnarray}
	\Delta(r)&=&a^2-2\mu r\,,\nonumber\\
	Q(r)&=&(1+B^2r^2)(a^2-2\mu r)\,.
\end{eqnarray}
Thus there is only one finite real root,
\begin{eqnarray}
	r_{\mathrm h}=\frac{a^2}{2\mu}\,,
\end{eqnarray}
whereas the other root is pushed to $r=\infty$ in the limit $A\to0$. The axial $\Omega=0$ locus is located at
\begin{equation}
	r_\Omega=-\mu\,,
\end{equation}
at $x=\pm1$.
In this sense, the surface $r=\infty$ should be regarded as the limiting
position of the second horizon. Radially outgoing null rays can still reach $r=\infty$ within finite affine parameter, since the behavior of the conformal factor $\Omega^2$ is unaffected.

In the following, we show that the compactification used in the generic case remains applicable. Introducing
\begin{eqnarray}
	y=-\frac{r_{\mathrm h}}{r}\,,
\end{eqnarray}
one finds
\begin{eqnarray}
	\widehat Q(y)
	&=&
	y^4 Q\left(-\frac{r_{\mathrm h}}{y}\right)\nonumber\\
	&=&
	y\left(y^2+B^2r_{\mathrm h}^2\right)
	\left(2\mu r_{\mathrm h}+a^2y\right)\,.
\end{eqnarray}
Therefore $\widehat Q$ has a simple zero at $y=0$, so $g_{yy}$ diverges at this surface. Next we show that this divergence is only a coordinate singularity, analogous to the standard singularity of Schwarzschild coordinates at a horizon.

\begin{figure}[htbp]
	\centering
	\resizebox{0.5\columnwidth}{!}{%
		\includegraphics{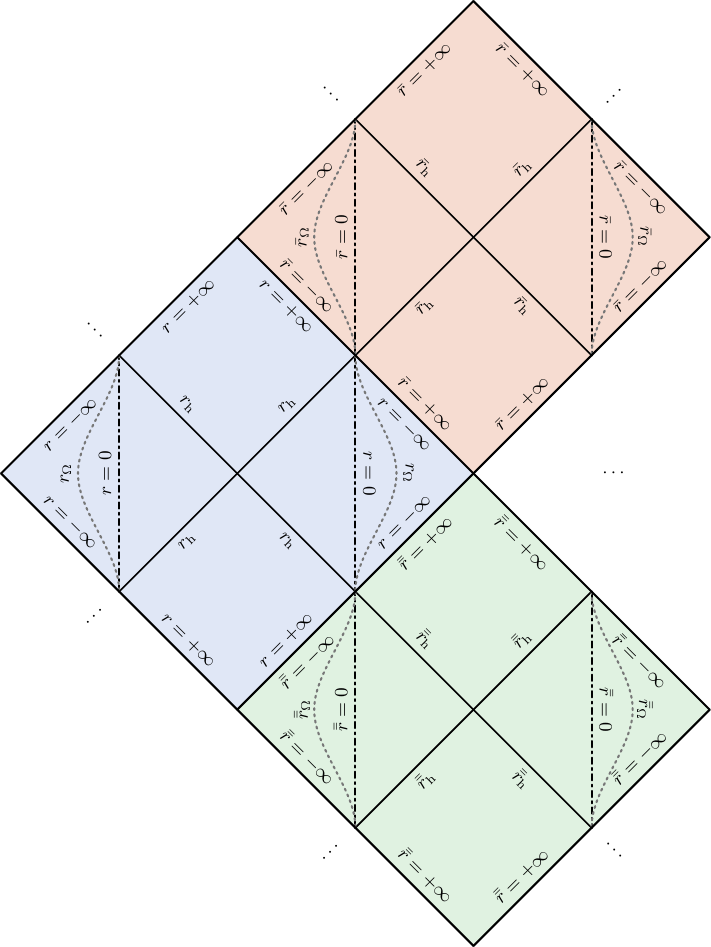}%
	}
	\caption{\justifying
		A portion of the Penrose diagram for the case $A=0$. The full diagram is obtained by extending this pattern indefinitely, with the ellipses indicating the extension directions. Here only two representative extensions are shown. Each color represents one universe, whose radial coordinate is denoted by $r$, $\bar r$, or $\bar{\bar r}$. %The hatched diamond marks a representative region where the constant-$r$ hypersurfaces are timelike.
	}
	\label{fig:vanishA-penrose}
\end{figure}

Actually, one can  introduce  null coordinates as
\begin{eqnarray}
	u=t-r_*\,,
	\quad
	v=t+r_*\,,
\end{eqnarray}
where
\begin{eqnarray}
	r_*=\frac{1}{2\kappa_{\mathrm h}}\log\left\lvert\frac{B(r-r_{\mathrm h})}{\sqrt{1+B^2r^2}}\right\rvert-\frac{1}{4\kappa_\infty}\log(1+B^2r^2)-\frac{2\mu^3B^3r_{\mathrm h}}{(1+\mu^2B^2)^2}\arctan(Br)\, .
\end{eqnarray}
Here
\begin{eqnarray}
	\kappa_\infty
	&=&
	\left.
	\frac{\mathrm{d}\widehat Q/\mathrm{d}y}
	{2r_{\mathrm h}\widehat\rho_0^2(y)}
	\right|_{y=0}
	=\mu B^2\, ,
	\nonumber\\
	\widehat\rho_0^2(y)
	&=&
	y^2\rho_0^2\left(-\frac{r_{\mathrm h}}{y}\right)
	=r_{\mathrm h}^2+a^2y^2\, .
\end{eqnarray}
Therefore the corresponding Kruskal-type coordinates adapted to the surface
$y=0$ are
\begin{eqnarray}
	U=-\exp(-\kappa_\infty u)\,,
	\quad
	V=\exp(\kappa_\infty v)\,.
\end{eqnarray}
They obey
\begin{eqnarray}
	UV&=&-\exp(2\kappa_\infty r_*)\sim y+\mathcal{O}(y^2)\,,\nonumber\\
	\frac{V}{U}&=&-\exp(2\kappa_\infty t)\,.
\end{eqnarray}
Crossing the null surface $UV=0$ changes the sign of $y$. The extended region with $y>0$ is therefore described by $r<0$, with $r\to-\infty$ as $y\to0^+$. Consequently, the $r=+\infty$ surface of one universe can be smoothly glued to the $\bar{r}=-\infty$ surface of a second universe with the same parameters. The resulting Penrose diagram is shown in
Fig.~\ref{fig:vanishA-penrose}.

\subsection{Sector with $A<0$}
% Representative interior point for the exact-curve construction: M=1,
% a=0.6, B=(b_-(0.6)+b_+(0.6))/2=1.290213689311.
In some regions of parameter space, $A$ can become negative. In this case, the two real roots of $Q$ have opposite signs, and the constant-$r$ hypersurfaces are timelike only in the region between them. These two roots should therefore no longer be interpreted as the inner and outer horizons. Instead, we denote them by $r_{\rm n}$ and $r_{\rm p}$, with $r_{\rm n}<0<r_{\rm p}$. In this sense, they more closely resemble a black-hole horizon and a cosmological horizon, respectively. The axial $\Omega=0$ locus remains at $r=r_\Omega$, given by Eq.~\eqref{rOm}, with $x=\pm1$, and one can show that $r_{\rm n}<r_\Omega<0$.

\begin{figure}[htbp]
	\centering
	\resizebox{0.5\columnwidth}{!}{%
		\includegraphics{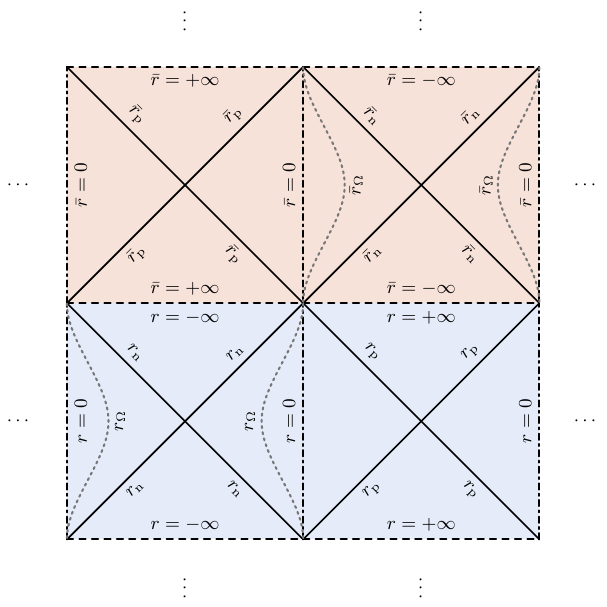}}
	\caption{\justifying Minimal tileable blocks for the $A<0$ Penrose diagram. The full Penrose diagram is obtained by gluing infinitely many copies of these blocks, with the ellipses indicating the infinite continuation of this construction. The blue and red strips denote the first and second universes, respectively. The dashed horizontal segments represent spacelike surfaces at $r=\pm\infty$ or $\bar r=\pm\infty$, along which the two universes are glued. The dashed vertical segments represent timelike surfaces at $r=0$ or $\bar r=0$, where ring singularities appear on the equatorial plane. %The hatched diamond marks a representative region in which the constant-$r$ surfaces are timelike.
	}
	\label{fig:negA-penrose}
\end{figure}

Nevertheless, null geodesics can still reach $r=\infty$ within finite affine parameter, so the same gluing construction remains applicable. Two universes can therefore be joined across the surface $r=\infty$, yielding the Penrose diagram shown in Fig.~\ref{fig:negA-penrose}. Since the surface $r=\infty$ is spacelike, regular data specified on it determine a unique continuation across this surface, at least within its domain of dependence~\cite{Choquet-Bruhat:1969ywq, Friedrich:2000qv, Rendall:2000pk}.

\subsection{Static limit with $a=0$}
When $a=0$, the radial function $Q(r)$ has only one real root,
\begin{eqnarray}
	r_{\mathrm h}=\frac{2M}{A}\,.
\end{eqnarray}
For $B\neq0$, the axial $\Omega=0$ locus is located at
\begin{equation}
	r_\Omega=-\frac{1}{B^2M}\,,
\end{equation}
which follows from $r_\Omega=-\mu/(1-A)$ with $\mu=M$ and $A=1-B^2M^2$.
The surface $r=0$ is now a curvature singularity and, unlike the ring singularity, cannot be bypassed through the non-equatorial plane. It is spacelike when approached from the $r>0$ side, but timelike when approached from the $r<0$ side. Radial null rays can still reach $r=\infty$ within finite affine parameter, so this surface is again not a boundary. The extension method used in the rotating case still works by introducing the coordinate $y=-r_{\mathrm h}/r$. Before the extension, the sign of $A=1-B^2M^2$ changes the structure of the spacetime dramatically. After extension, however, these cases are naturally unified into the same causal structure, differing only in the causal characters of the $r=\infty$ surfaces.

\begin{figure*}[htbp]
	\centering

	% ============================================================
	% A > 0
	% ============================================================
	\begin{subfigure}[t]{0.31\textwidth}
		\centering
		\resizebox{\linewidth}{!}{%
			\includegraphics{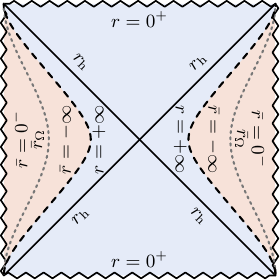}}
		\caption{$A>0$}
		\label{fig:static-a0-penrose-A-positive}
	\end{subfigure}
	\hfill
	% ============================================================
	% A = 0
	% ============================================================
	\begin{subfigure}[t]{0.31\textwidth}
		\centering
		\resizebox{\linewidth}{!}{%
			\includegraphics{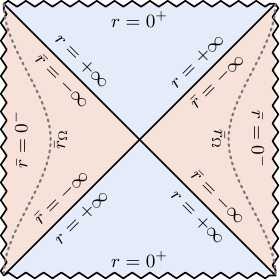}}
		\caption{$A=0$}
		\label{fig:static-a0-penrose-A-zero}
	\end{subfigure}
	\hfill
	% ============================================================
	% A < 0
	% ============================================================
	\begin{subfigure}[t]{0.31\textwidth}
		\centering
		\resizebox{\linewidth}{!}{%
			\includegraphics{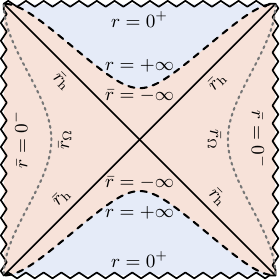}}
		\caption{$A<0$}
		\label{fig:static-a0-penrose-A-negative}
	\end{subfigure}

	\caption{\justifying Penrose diagrams for the static $a=0$ limit. Blue shading denotes the $r>0$ regions, while red shading denotes the $\bar r<0$ regions.}
	\label{fig:static-a0-penrose}
\end{figure*}

For $A>0$, the root $r_{\mathrm h}$ is positive and the surface $r=\infty$ is timelike. The positive-radius exterior is glued through this timelike surface to a negative-radius region. The corresponding Penrose diagram is shown in Fig.~\ref{fig:static-a0-penrose}(a).

For $A=0$, the finite root is pushed to $r=\infty$, which therefore becomes a null surface. In this case one may instead use $y=-M/r$. It can be proved that the apparent singularity of the metric in the $y$ coordinate is merely a coordinate singularity. The resulting Penrose diagram is shown in Fig.~\ref{fig:static-a0-penrose}(b).

For $A<0$, the root $r_{\mathrm h}$ lies at negative radius. There is no positive-radius exterior bounded by a positive real root. The surface $r=+\infty$ is spacelike and is glued to $\bar r=-\infty$ of a negative-radius region. This region contains a horizon at $\bar r=\bar r_{\mathrm h}$ and terminates at the timelike singularity $\bar r=0^-$. The corresponding diagram is shown in Fig.~\ref{fig:static-a0-penrose}(c).

\subsection{Massless limit and global Bertotti-Robinson geometry}
The massless limit of the KBR family is known to be globally equivalent to the Bertotti-Robinson spacetime, namely the direct product $\mathrm{AdS}_2\times S^2$~\cite{Podolsky:2025tle,Griffiths:2009dfa,Ottewill:2012mq}. Nevertheless, it is natural to ask whether the same approach used in the $M\neq0$ case, namely
\begin{eqnarray}
	y=-\frac{1}{r},
\end{eqnarray}
can extend the spacetime across the surface $r=+\infty$.

At the would-be bridge surface $y=0$, one finds
\begin{eqnarray}
	\widehat{\Omega}^{2}(0,x)=B^2(1-x^2)\,.
\end{eqnarray}
Therefore, for every fixed non-axial direction $|x|<1$,
\begin{eqnarray}
	\widehat{\Omega}^{2}(0,x)>0,
\end{eqnarray}
so the hypersurface $y=0$ is perfectly regular and timelike. In this sense, the coordinate $y=-1/r$ does provide a local extension across $r=+\infty$.

However, this extension is not global. Indeed, at the symmetry axis, $x=\pm1$, one has
\begin{eqnarray}
	\widehat{\Omega}^{2}(0,\pm1)=0\,.
\end{eqnarray}
Since the metric carries the overall factor $\widehat{\Omega}^{-2}$, the $y$-chart degenerates at the two axis points of the $y=0$ surface. Thus the transformation $y=-1/r$ furnishes only a local extension for $|x|<1$.

This behavior contrasts with the $M\neq0$ case, where $\widehat{\Omega}^{2}(0,\pm1)$ remains strictly positive. Consequently, the axial obstruction is absent when $M\neq0$, and the same inverse-radius compactification yields a smooth bridge across $r=\infty$.

The local $y$ continuation should not be interpreted as producing distinct BR universes. In the massless limit the KBR metric is exactly the Bertotti-Robinson spacetime $\mathrm{AdS}_2\times S^2$, and the known transformation to global BR coordinates covers the repeated massless KBR patches within a single global geometry~\cite{Podolsky:2025tle,Griffiths:2009dfa,Ottewill:2012mq}. In that description, the original surface $r=\infty$ lies in the BR bulk rather than at its conformal boundary. The true asymptotic boundary becomes manifest only in the global BR coordinates. This limit therefore illustrates that repeated use of a local $1/r$ continuation need not by itself constitute a maximal global description.

\section{Wave Scattering on the Extended Spacetime}\label{sec-waves}

Motivated by the global causal structure, we use a test massless scalar field to probe the spectral properties of the extended geometry. The analysis concerns a fixed two-universe scattering segment and should not be identified with a globally well-posed Cauchy evolution of the full spacetime.

\subsection{Separated equations and scattering problem}
We study a test massless Klein-Gordon field obeying
\begin{eqnarray}\label{KG}
	\square\Phi=0\,.
\end{eqnarray}
The separability of the KG equation~\eqref{KG} in the KBR geometry was established in~\cite{Gray:2025lwy}. Here we provide a complementary discussion. The complicated form of the overall conformal factor $\Omega$ makes its separability less clear. Consider the conformal KG equation,
\begin{eqnarray}
	\left(\square-\frac{1}{6}\mathcal{R}\right)\Phi=0\,,
\end{eqnarray}
which is equivalent to Eq.~(\ref{KG}) since the Ricci scalar $\mathcal{R}$ vanishes, thanks to the tracelessness of the energy-momentum tensor of the Maxwell field. Introducing the Yamabe (conformal Laplacian) operator $Y_g\coloneqq\square_g-\frac{1}{6}\mathcal{R}[g]$, we note that it is conformally invariant in the sense that
\begin{eqnarray}
	Y_g(\psi)=\Omega^3Y_{\tilde g}(\tilde\psi)\,,
\end{eqnarray}
where
\begin{eqnarray}
	\tilde g_{ab}=\Omega^2 g_{ab}\,,\quad \tilde\psi=\Omega^{-1}\psi\,.
\end{eqnarray}
The Ricci scalar $\mathcal{R}[\tilde g]$ separates into $r$- and $x$-dependent parts, so the conformally related wave equation has a manifestly separable structure. The conformal covariance then carries this separability back to the physical metric, with the factor of $\Omega$ in the scalar-field ansatz accounting precisely for the conformal weight. Accordingly, we adopt the ansatz
\begin{eqnarray}
	\Phi(t,r,x,\varphi)=\Omega\mathrm{e}^{-\mathrm{i}(\omega t-m\varphi)}\frac{\Phi^{(r)}(r)}{r}\Phi^{(x)}(x)\,,
\end{eqnarray}
Eq.~\eqref{KG} separates into radial and angular parts:
\begin{eqnarray}
	\frac{\mathrm{d}}{\mathrm{d}r}\left(\frac{Q}{r^2}
	\frac{\mathrm{d}\Phi^{(r)}}{\mathrm{d}r}\right)
	+\frac{1}{r^2}\Bigg[\frac{\left(\omega\rho_0^2-a m/C\right)^2}{Q}-\frac{2}{r^2}(\mu r-a^2)-\lambda\Bigg]\Phi^{(r)}=0\,,\label{eqr}\\
	\frac{\mathrm{d}}{\mathrm{d}x}\left(P\Delta_x
	\frac{\mathrm{d}\Phi^{(x)}}{\mathrm{d}x}\right)
	-\Bigg[\frac{\left(a\omega\Delta_x-m/C\right)^2}{P\Delta_x}+2(P-1)-\lambda\Bigg]\Phi^{(x)}=0\,.\label{eqx}
\end{eqnarray}
Here the separation constant $\lambda$ reduces to the Kerr one $\lambdabar=A_{\ell m}(a\omega)+a^2\omega^2-2am\omega$ as $B$ vanishes, and the azimuthal number $m\in\mathbb{Z}$ as the conicity factor $C$ has already been explicitly incorporated in the metric. Although there is an explicit $r^{-1}$ factor in the ansatz for $\Phi$, the field remains regular at $r=0$. Indeed, the radial equation~\eqref{eqr} implies that $\Phi^{(r)}=\mathcal{O}(r)$ as $r\to0$, so the combination $\Phi^{(r)}/r$ stays finite.

We impose an ingoing boundary condition at the left boundary and an outgoing boundary condition at the right boundary.
\begin{eqnarray}
	\Phi^{(r)}&\sim& \mathrm{e}^{-\mathrm{i}(\omega-m\varpi_{\text{o}})r_*}\,,\quad r\to r_{\text{o}}\,, \nonumber\\
	\bar{\Phi}^{(\bar{r})}&\sim& \mathrm{e}^{\mathrm{i}(\omega-m\bar{\varpi}_{\text{i}})\bar{r}_*}\,,\quad\;\;\; \bar{r}\to \bar{r}_{\text{i}}\,.
\end{eqnarray}
The outgoing condition at the right boundary describes waves leaving the scattering region toward the white-hole-like region of the second universe. 

\begin{figure*}[htbp]
		\centering
		\includegraphics{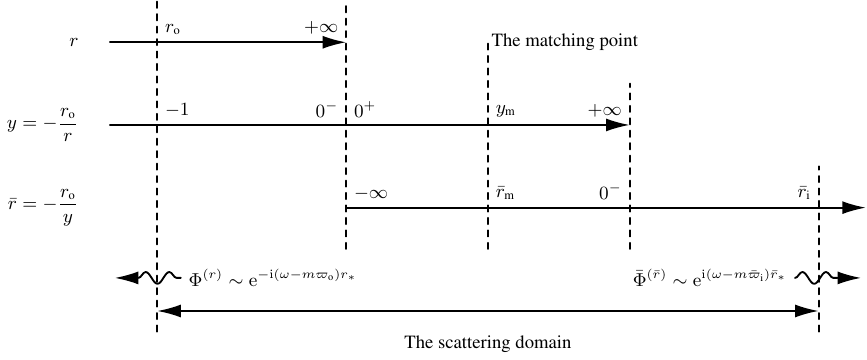}
		\caption{\justifying A sketch of the coordinate patches, the scattering domain, and the boundary conditions.}
		\label{fig:coorpatch}
	\end{figure*}

These boundary conditions define a two-universe scattering segment from the outer horizon $r=r_{\text{o}}$ of the first region to the inner horizon $\bar r=\bar r_{\text{i}}$ of the neighboring region. The tortoise coordinate is normalized by $r_*=0$ at the gluing surface and runs from $-\infty$ at the former endpoint to $+\infty$ at the latter. The coordinate patches and boundary conditions are shown schematically in Fig.~\ref{fig:coorpatch}.

In the $y$ coordinate, the radial Eq.~\eqref{eqr} becomes
\begin{eqnarray}\label{eqy}
	\frac{\mathrm{d}}{\mathrm{d}y}\left(\frac{\widehat{Q}}{r_{\text{o}}^2y^2}
	\frac{\mathrm{d}\Phi^{(y)}}{\mathrm{d}y}\right)
	+\frac{1}{y^2}\left[\frac{\left(\omega\widehat{\rho}_0^2-a m y^2/C\right)^2}{\widehat{Q}}+\frac{2B^2r_{\text{o}}}{y^2}(Ar_{\text{o}}+\mu y)-\lambda\right]
	\Phi^{(y)}=0\,,
\end{eqnarray}
where $$\Phi^{(y)}(y)=y\Phi^{(r)}(-r_{\text{o}}/y)\, .$$ Its Schr\"odinger-like form is
\begin{eqnarray}\label{eqystar}
	\frac{\mathrm{d}^2\Phi^{(y_*)}}{\mathrm{d}r_*^2}
	+\frac{\widehat{Q}}{\widehat{\rho}_0^4}\left[
	\frac{\left(\omega\widehat{\rho}_0^2-a m y^2/C\right)^2}{\widehat{Q}}
	+y\widehat{\rho}_0\frac{\mathrm{d}}{\mathrm{d}y}
	\left(\frac{\widehat{Q}}{y^2\widehat{\rho}_0^3}\right)	+\frac{2B^2r_{\text{o}}}{y^2}(Ar_{\text{o}}+\mu y)-\lambda
	\right]\Phi^{(y_*)}=0\,,
\end{eqnarray}
where $$\Phi^{(r_*)}=\rho_0\Phi^{(r)}/r\, ,\qquad  \Phi^{(y_*)}=\widehat{\rho}_0\Phi^{(y)}/y\, ,$$ and $\mathrm{d}r_*/\mathrm{d}y=r_{\text{o}}\widehat{\rho}_0^2/\widehat{Q}$.
A Liouville transformation casts the radial equation into a Schr\"odinger-like form
\begin{equation}
	\frac{\mathrm{d}^2\Phi^{(r_*)}}{\mathrm{d}r_*^2}+V\Phi^{(r_*)}=0\,,\label{eqrstar}
    \end{equation}
and the effective potential is given by
    \begin{equation}
	V=\frac{Q}{\rho_0^4}\left[\frac{\left(\omega\rho_0^2-a m/C\right)^2}{Q}
		+a^2r\rho_0\frac{\mathrm{d}}{\mathrm{d}r}\left(\frac{Q}{r^2\rho_0^3}\right)-\frac{2}{r^2}(\mu r-a^2)-\lambda\right]\,.
\end{equation}

\subsection{Effective Potential and Trapping}\label{sec-potential}
We first examine the behavior of this effective radial potential $V$. Consider a monochromatic wave with real frequency $\omega$, for which the separation constant $\lambda$ is also real. In the WKB regime, the solution is accurately described by~\cite{Comins, Cardoso:2007az, Patrick:2018orp}
\begin{eqnarray}
	\Phi^{(r_*)} \sim A_0 \exp\left[\mathrm{i}\int k(r_*, \omega)\mathrm{d}r_*\right] \,,
\end{eqnarray}
where $A_0$ is a slowly varying amplitude and $k(r_*, \omega)$ is the local wave number satisfying the Hamilton-Jacobi equation
\begin{eqnarray}
	k^2(r_*, \omega) = V(r_*, \omega)=\mathcal{F}(r_*, \omega)\prod_{i=1}^N(\omega-\omega_i(r_*))\,,
\end{eqnarray}
where $\omega_i$ are real, $\mathcal F$ is a smooth positive function, and $N$ counts the real roots at fixed $r_*$. The number of real roots can exceed two because a branch may fold back on itself. In the $m=0$ sector there are also intervals with no real roots, where the relevant pair becomes purely imaginary. Positive $k^2$ corresponds to an oscillatory region, negative $k^2$ to an evanescent region, and a turning point occurs where $k^2=0$.

\begin{figure*}[htbp]
\centering
\begin{subfigure}{0.3\textwidth}
\centering
\includegraphics[width=\linewidth]{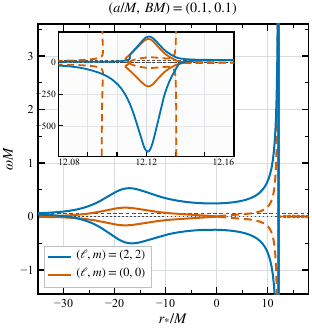}
\end{subfigure}\hfill
\begin{subfigure}{0.3\textwidth}
\centering
\includegraphics[width=\linewidth]{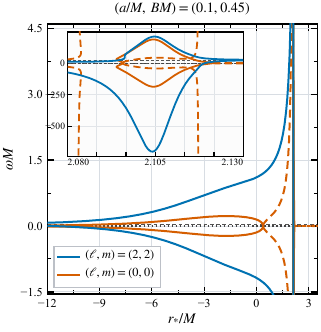}
\end{subfigure}\hfill
\begin{subfigure}{0.3\textwidth}
\centering
\includegraphics[width=\linewidth]{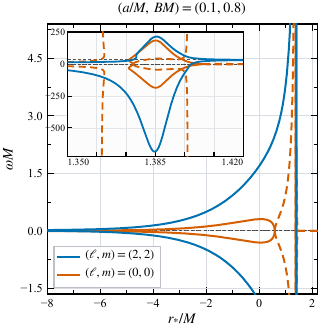}
\end{subfigure}\\

\begin{subfigure}{0.3\textwidth}
\centering
\includegraphics[width=\linewidth]{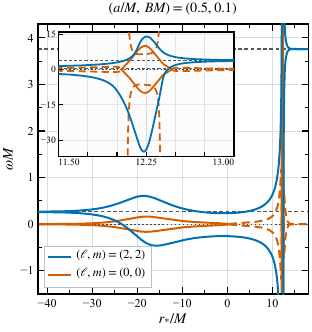}
\end{subfigure}\hfill
\begin{subfigure}{0.3\textwidth}
\centering
\includegraphics[width=\linewidth]{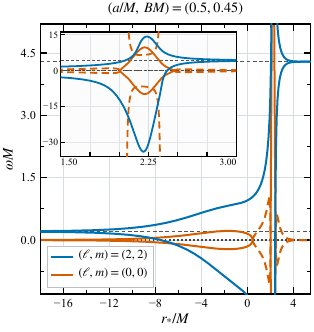}
\end{subfigure}\hfill
\begin{subfigure}{0.3\textwidth}
\centering
\includegraphics[width=\linewidth]{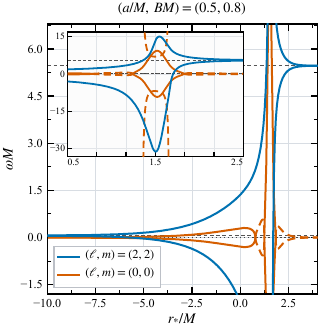}
\end{subfigure}\\

\begin{subfigure}{0.3\textwidth}
\centering
\includegraphics[width=\linewidth]{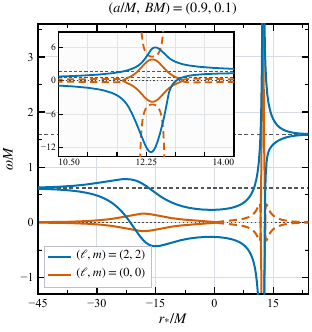}
\end{subfigure}\hfill
\begin{subfigure}{0.3\textwidth}
\centering
\includegraphics[width=\linewidth]{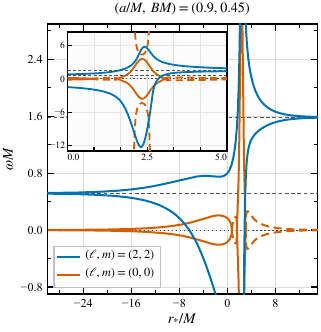}
\end{subfigure}\hfill
\begin{subfigure}{0.3\textwidth}
\centering
\includegraphics[width=\linewidth]{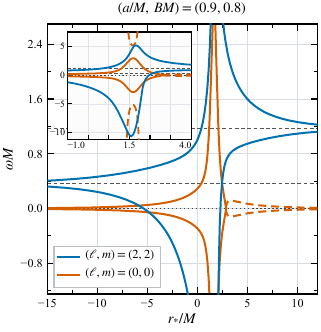}
\end{subfigure}
\caption{\justifying The roots $\omega_i(r_*)$ satisfying $V(r_*;\omega_i,\lambda(\omega_i))=0$ for representative parameter choices. Rows correspond to $a/M=0.1$, $0.5$, and $0.9$, while columns correspond to $BM=0.1$, $0.45$, and $0.8$. Solid curves denote real roots and dashed curves denote the imaginary parts of purely imaginary roots. The horizontal dashed lines indicate the asymptotic horizon frequencies, while the vertical dashed lines mark the equatorial locations where $g_{\varphi\varphi}=0$.}
\label{fig:po}
\end{figure*}

The separation constant $\lambda$ is determined by the angular equation and depends nontrivially on $\omega$. For each real frequency we therefore first solve the angular problem, insert $\lambda(\omega)$ into the radial potential, and then determine the turning-point roots. Representative roots are shown in Fig.~\ref{fig:po}. They approach $m\varpi_{\mathrm o}$ and $m\bar\varpi_{\mathrm i}$ as $r_*\to-\infty$ and $r_*\to+\infty$, where
\begin{equation}
\varpi_j=\left.\frac{a}{C\rho_0^2}\right|_{r=r_j}
\end{equation}
denotes the angular velocity associated with the corresponding Killing horizon. Typically the potential contains two barriers, one in each KBR region, which form a cavity across the gluing surface. Only for sufficiently large $BM$ or sufficiently small $a/M$ does the first-region barrier disappear. The wormhole-like global structure is therefore accompanied by a characteristic trapping geometry for waves.

The real-frequency root structure gives a direct view of propagation and trapping. Positive $V$ corresponds to an oscillatory WKB region, while negative $V$ gives an evanescent region. Depending on the parameters, a branch can fold so that as many as four real roots appear at a fixed $r_*$; in the axisymmetric case some roots instead move onto the imaginary-frequency axis. Typically, the potential contains two barriers, one in each universe, and a cavity spans the gluing surface. The first-universe barrier disappears for sufficiently large $BM$ or sufficiently small $a/M$, so the cavity occupies a broader parameter region than the unstable nonaxisymmetric modes discussed below and provides the natural trapping region for long-lived modes and echo-like responses.

\subsection{Axisymmetric Growing Modes and Chronology Violation}\label{sec-axisymmetric}
As in the Kerr spacetime, the extended KBR spacetime contains closed timelike curves in the $r<0$ region, where the associated chronology violation is closely related to the existence of exponentially growing modes in the Kerr case~\cite{Dotti:2011ix}. For axisymmetric modes with $m=0$, we consider purely imaginary frequencies of the form $\omega=i\gamma/a$ with real positive $\gamma$ and, following~\cite{Dotti:2011ix}, show that QNMs of this form exist. In this case, the radial QNM boundary conditions reduce to exponential decay and both the radial and angular equations can be cast as real self-adjoint eigenvalue problems, with the corresponding eigenvalues denoted as $E_{\text{r}}$ and $E_{\text{a}}$, respectively:
\begin{align}
\mathcal{R}\Phi^{(r_*)}&=-E_{\text{r}}\frac{Q}{\rho_0^4}\Phi^{(r_*)}\,,&
\mathcal{R}&=\mathcal{R}_0+\gamma^{2}\mathcal{R}_2\,,\\
\mathcal{A}\Phi^{(x)}&=E_{\text{a}}\Phi^{(x)}\,,&
\mathcal{A}&=\mathcal{A}_0+\gamma^2\mathcal{A}_2\,,
\end{align}
where
\begin{align}
\mathcal{R}_0&=\mathcal{W}^\dagger\mathcal{W}+\frac{B^2\Delta^2}{\rho_0^4}\,,\nonumber\\
\mathcal{W}&=\frac{\mathrm{d}}{\mathrm{d}r_*}-\frac{(1-B^2a^2)r\Delta}{\rho_0^4}\,,\nonumber\\
\mathcal{W}^\dagger&=-\frac{\mathrm{d}}{\mathrm{d}r_*}-\frac{(1-B^2a^2)r\Delta}{\rho_0^4}\,,\nonumber\\
\mathcal{R}_2&=\frac{1}{a^2}-\frac{Q}{\rho_0^4}\,,\nonumber\\
\mathcal{A}_0&=-\frac{\mathrm{d}}{\mathrm{d}x}\left(P\Delta_x\frac{\mathrm{d}}{\mathrm{d}x}\right)
+2(P-1)\,,\nonumber\\
\mathcal{A}_2&=\frac{x^2}{CP}\,.\nonumber
\end{align}
Since $Q/\rho_0^4>0$ in the interior of the scattering domain, the radial equation defines a generalized Sturm--Liouville problem with positive weight. A QNM is obtained when $$E_{\text{r}}=E_{\text{a}}=\lambda+\gamma^2\, .$$We then compare the radial and angular eigenvalues in the $\gamma\to0^+$ and $\gamma\to+\infty$ limits and show that they necessarily intersect for every fixed $\ell$.

We first focus on the radial sector. For sufficiently large $\gamma$, first note that near $\bar r=0$, the function $\mathcal{R}_2$ possesses a negative region that coincides precisely with the $g_{\varphi\varphi}<0$ region on the equatorial plane and is always present. Take the standard $L^2$ inner product and norm with respect to the tortoise coordinate, and choose a fixed trial function $\psi$ with compact support entirely contained in the region where $\mathcal{R}_2<0$, normalized such that $\langle\psi,Q\psi/\rho_0^4\rangle=1$. By the Rayleigh--Ritz variational principle, the fundamental radial eigenvalue satisfies
\begin{eqnarray}
-E_{\text{r}}^{(0)}(\gamma)
\leqslant
\langle\psi,\mathcal{R}_0\psi\rangle
+\gamma^2\langle\psi,\mathcal{R}_2\psi\rangle.
\end{eqnarray}
Therefore,
$$E_{\text{r}}^{(0)}(\gamma)\geqslant c\gamma^2+\mathcal{O}(\gamma^0)\, ,$$
where $c=-\langle\psi,\mathcal{R}_2\psi\rangle>0$.

For small $\gamma$, let $\psi$ instead be an eigenfunction corresponding to $E_{\text{r}}^{(0)}(\gamma)$, still normalized such that $\langle\psi,Q\psi/\rho_0^4\rangle=1$. One can obtain that 
\begin{eqnarray}
\gamma^2-E_{\text{r}}^{(0)}(\gamma)&=&\left\lVert\mathcal{W}\psi\right\rVert^2+\left\langle\psi,\frac{B^2\Delta^2}{\rho_0^4}\psi\right\rangle+\frac{\gamma^2}{a^2}\left\lVert\psi\right\rVert^2>0\,.
\end{eqnarray}
Hence,
\begin{eqnarray}
E_{\text{r}}^{(0)}(\gamma)<\gamma^2
\end{eqnarray}
for every $\gamma>0$, and therefore $\lim_{\gamma\to0^+}E_{\text{r}}^{(0)}(\gamma)\leq0$.

We next consider the angular eigenvalue. For large $\gamma$, we introduce the rescaled coordinate $\zeta=\sqrt{\gamma}x$. The angular operator reduces at leading order to
\begin{eqnarray}
\mathcal{A}=\gamma\left(-\frac{\mathrm{d}^2}{\mathrm{d}\zeta^2}+\frac{\zeta^2}{C}\right)+\mathcal{O}(\gamma^0)\,.
\end{eqnarray}
Therefore,
\begin{eqnarray}
E_{\text{a}}^{(\ell)}(\gamma)=\frac{2\ell+1}{\sqrt{C}}\gamma+\mathcal{O}(\gamma^0)\,.\label{ang eig large gam}
\end{eqnarray}
Thus, the angular eigenvalue grows only linearly with $\gamma$, whereas the fundamental radial eigenvalue grows at least quadratically. Consequently, for sufficiently large $\gamma$, we have $$E_{\text{r}}^{(0)}(\gamma)>E_{\text{a}}^{(\ell)}(\gamma)\, .$$

For small $\gamma$, as the angular domain and boundary conditions are independent of $\gamma$, and $\mathcal{A}_2$ is bounded, the angular eigenvalues depend continuously on $\gamma$, in particular at $\gamma=0$. Since $\mathcal{A}_2\geq0$, the min--max principle gives
\begin{eqnarray}
E_{\text{a}}^{(\ell)}(\gamma)\geq E_{\text{a}}^{(\ell)}(0).
\end{eqnarray}
Moreover, 
\begin{eqnarray}
\left\langle\psi,\mathcal{A}_0\psi\right\rangle=\int_{-1}^{1}\left[P\Delta_x\left|\frac{\mathrm{d}\psi}{\mathrm{d}x}\right|^2+2(P-1)|\psi|^2\right]\mathrm{d}x>0\,.
\end{eqnarray}
Thus, $E_{\text{a}}^{(\ell)}(0)>0$ for every $\ell$.
One can choose a sufficiently small $\gamma>0$ such that
\begin{eqnarray}
E_{\text{r}}^{(0)}(\gamma)<\gamma^2<E_{\text{a}}^{(\ell)}(0)\leq E_{\text{a}}^{(\ell)}(\gamma)\,.
\end{eqnarray}
Note that the inequality reverses between the two limiting cases. Continuity of the radial and angular eigenvalues with respect to $\gamma>0$ therefore guarantees the existence of some $\gamma>0$ for which
\begin{eqnarray}
E_{\text{a}}^{(\ell)}(\gamma)=E_{\text{r}}^{(0)}(\gamma).
\end{eqnarray}
Therefore, for every $\ell$, there exists an exponentially growing mode with $m=0$.

The purely imaginary unstable mode found above can be given a complementary interpretation in terms of the imaginary-frequency roots of the effective potential. The large-$\gamma$ angular asymptotic Eq.~\eqref{ang eig large gam} gives $\lambda=-\gamma^2+O(\gamma)$, so that
$$V(r_*;i\gamma/a)=-\gamma^2\mathcal{R}_2+O(\gamma)\, .$$Hence an imaginary root of $V=0$ can diverge as $\mathcal{R}_2\to0^-$, explaining the vertical asymptotes of the imaginary-root branches in Fig.~\ref{fig:po}. Between these asymptotes, $V>0$ for sufficiently large $\gamma$, producing an oscillatory region for imaginary-frequency perturbations bounded by regions with $V<0$. This bound-state-like configuration can support a discrete mode with exponentially decaying spatial tails. The existence of such a mode then implies exponential growth in time. Remarkably, this is precisely the same $\mathcal{R}_2<0$ region that enters the analytical existence proof above and, in the present geometry, coincides with the equatorial chronology-violating region containing closed timelike curves. The divergent imaginary roots therefore provide a local WKB picture of the same chronology-violating mechanism underlying the purely imaginary unstable mode.

In addition to the analytical argument above, we have also numerically confirmed the existence of purely imaginary QNMs with positive imaginary parts for the $\ell=m=0$ mode for several representative values of $a$ and $B$ using the method developed in the next section. The resulting modes are listed in Table~\ref{tab:pure_imag_qnms}.

\begin{table*}[t]
\centering
\begin{tabular}{cccc}
\hline\hline
$a$ & $B$ & $\omega$ & $\lambda$ \\
\hline
$0.1$ & $0.1$ & $43.0972929741\mathrm{i}$ & $-15.0624926074$ \\
$0.1$ & $0.45$ & $43.8612217720\mathrm{i}$ & $-15.2569152710$ \\
$0.1$ & $0.8$  & $45.4071191739\mathrm{i}$ & $-15.6495270825$ \\
$0.5$ & $0.1$  & $10.8336271656\mathrm{i}$ & $-24.7034622560$ \\
$0.5$ & $0.45$ & $11.0995020146\mathrm{i}$ & $-25.6437607415$ \\
$0.5$ & $0.8$  & $11.8539295360\mathrm{i}$ & $-28.7314720041$ \\
$0.9$ & $0.1$  & $8.3428859316\mathrm{i}$  & $-49.6441264923$ \\
$0.9$ & $0.45$ & $8.4728532894\mathrm{i}$  & $-51.1650307908$ \\
$0.9$ & $0.8$  & $10.0608511148\mathrm{i}$ & $-73.5150281907$ \\
\hline\hline
\end{tabular}
\caption{Numerically obtained purely imaginary unstable QNMs with $\ell=m=0$.}
\label{tab:pure_imag_qnms}
\end{table*}

However, the presence of chronology violation also indicates that the scattering region is not globally hyperbolic, so a globally well-defined Cauchy initial-value problem for perturbations is unavailable. Consequently, the existence of exponentially growing $m=0$ modes does not by itself establish a dynamical instability of the spacetime in the usual sense of evolution from generic initial data. To treat generic complex frequencies and compute the nonaxisymmetric spectrum, we next develop the exact Heun construction of the QNMs.

\subsection{Exact Heun Construction of the Quasinormal Modes}\label{sec-heun}
For generic complex frequency and nonzero $m$, the radial and angular problems no longer admit the real self-adjoint formulation used in Sec.~\ref{sec-axisymmetric}. Both equations can nevertheless be reduced to general Heun form~\cite{Suzuki:1998vy, Fiziev:2011mm, Hatsuda:2020sbn, Motohashi:2021zyv, Wu:2025wbp}. Here, we provide further details on the Heun functions and the matching procedure.

\subsubsection{Angular Heun problem}\label{sec-an}
We first concentrate on the angular equation~\eqref{eqx} which can be recast to a Heun equation
\begin{eqnarray}\label{heun}
	\frac{\mathrm{d}^2S}{\mathrm{d}z_x^2}
	+\left(\frac{\gamma_x}{z_x}+\frac{\delta_x}{z_x-1}
	+\frac{\epsilon_x}{z_x-z_{ax}}\right)
	\frac{\mathrm{d}S}{\mathrm{d}z_x}
	+\frac{\alpha_x\beta_x z_x-q_x}{z_x(z_x-1)(z_x-z_{ax})}S=0\,,
\end{eqnarray}
where
\begin{align*}
	z_x      & =1-x^{2}\,,\\
	S        & =z_x^{\frac{1-\gamma_x}{2}}(z_x-z_{ax})^{\frac{1-\epsilon_x}{2}}\Phi^{(x)}\,, \\
	z_{ax}   & =1+\frac{1}{p}\,,\\
	p        & =\frac{P-1}{x^{2}}=\frac{1}{C}-1\,,                                           \\
	\gamma_x & =1+\lvert m\rvert\,,\\
	\delta_x & =\frac{1}{2}\,,\\
	\epsilon_x& =1+\mathrm{i}\frac{a\omega-mp}{\sqrt{p}}\,,                          \\
	\alpha_x & =\frac{1}{2}+\frac{\gamma_x-1}{2}+\frac{\epsilon_x-1}{2}\,,\\
	\beta_x  & =1+\frac{\gamma_x-1}{2}+\frac{\epsilon_x-1}{2}\,,                       \\
	q_x      & =\frac{1}{2}+\frac{1}{4p}\Big[
	-\lambda-2a\omega m+\lvert m\rvert(\lvert m\rvert+1)(1+3p)
		+2\mathrm{i}(a\omega-mp)(\lvert m\rvert+1)\sqrt{p}\Big]\,,
\end{align*}
and in which $z_{ax}$ corresponds to the root of the function $P$. Note that the Fuchs relation holds:
\begin{eqnarray}
	\alpha_x+\beta_x-\gamma_x-\delta_x-\epsilon_x+1=0\,.\nonumber
\end{eqnarray}
Denote the local solution of Eq.~\eqref{heun} that is regular at $z_x=0$ as $H\ell(z_{ax},q_x;\alpha_x,\beta_x,\gamma_x,\delta_x;z_x)$, which is normalized as $H\ell(z_{ax},q_x;\alpha_x,\beta_x,\gamma_x,\delta_x;0)=1$. The physical local solution at $z_x=0$ is
\begin{eqnarray}\label{S0}
	S_0(z_x)=H\ell(z_{ax},q_x;\alpha_x,\beta_x,\gamma_x,\delta_x;z_x)\,,
\end{eqnarray}
and the two local solutions at $z_x=1$ are
\begin{eqnarray}
	S_{11}(z_x)&=&H\ell(1-z_{ax},\alpha_x\beta_x-q_x;
	\alpha_x,\beta_x,\delta_x,\gamma_x;1-z_x)\,,\nonumber\\
	S_{12}(z_x)&=&(1-z_x)^{1-\delta_x}
	H\ell(1-z_{ax},q_{x1};\alpha_x-\delta_x+1,\beta_x-\delta_x+1,
	2-\delta_x,\gamma_x;1-z_x)\nonumber\\
	&=&|x|H\ell(1-z_{ax},q_{x1};\alpha_x-\delta_x+1,\beta_x-\delta_x+1,	2-\delta_x,\gamma_x;1-z_x)\,,
\end{eqnarray}
where
\begin{eqnarray}
	q_{x1}&=&\alpha_x\beta_x-q_x+(1-\delta_x)(\gamma_x(1-z_{ax})+\epsilon_x)\,.\nonumber
\end{eqnarray}
They are related by
\begin{eqnarray}
	S_0(z_x)=C_1S_{11}(z_x)+C_2S_{12}(z_x)\,,
\end{eqnarray}
where the coefficients $C_i$ can be calculated via
\begin{eqnarray}
	C_1=\frac{W(S_0,S_{12})}{W(S_{11},S_{12})}\,,\quad C_2=\frac{W(S_0,S_{11})}{W(S_{12},S_{11})}\,,
\end{eqnarray}
in which $W(f,g)\equiv fg'-f'g$ is the Wronskian.

After recasting the angular equation~\eqref{eqx} into Heun equation~\eqref{heun}, and choosing the regular local Heun function~\eqref{S0} at $z=0$, we obtain a solution that automatically satisfies the boundary conditions at $x=-1$ and $x=1$ for any $\lambda$. In fact, it is the constraint on $C_1$ and $C_2$ at $x=0$ ($z=1$) that selects a discrete set of admissible $\lambda$. Since the original angular equation in $x$ [Eq.~\eqref{eqx}] is symmetric under $x\rightarrow-x$, one may consider only the odd and even parity solutions, which correspond to $C_1=0$ and $C_2=0$, respectively. An alternative viewpoint is to require the physical solution $\Phi^{(x)}(x)$ to be smooth at $x=0$. Since the map $z_x=1-x^2$ is two-to-one on $x\in[-1,1]$, the intervals $x\in[-1,0]$ and $x\in[0,1]$ correspond to two branches of the same solution $S_0(z_x)$, which may differ only by independent overall normalizations $A_-$ and $A_+$. The prefactor relating $\Phi^{(x)}$ and $S$ depends only on $x^2$ and has vanishing derivative at $x=0$, so it suffices to impose smoothness directly on $S$. Since $S_{11}=1+\mathcal{O}(x^2)$ and $S_{12}=|x|[1+\mathcal{O}(x^2)]$ near $x=0$, the two branches behave as
\begin{eqnarray}
S^{(+)}(x)&=&A_+\left[C_1+C_2x+\mathcal{O}(x^2)\right]\,,\qquad x\to0^+\,,\nonumber\\
S^{(-)}(x)&=&A_-\left[C_1-C_2x+\mathcal{O}(x^2)\right]\,,\qquad x\to0^-\,.
\end{eqnarray}
Continuity of the solution and its first derivative at $x=0$ therefore requires
\begin{eqnarray}
(A_+-A_-)C_1=0\,,\qquad (A_++A_-)C_2=0\,.
\end{eqnarray}
For a nontrivial global solution, these conditions are compatible only if
\begin{eqnarray}\label{anCon}
C_1=0\quad \text{or}\quad C_2=0\,.
\end{eqnarray}
Thus the smooth-matching condition reproduces the same discrete angular spectrum obtained from the parity decomposition.

In practice, $C_1$ and $C_2$ are evaluated at $z_x=1/2$. Singular points of Eq.~\eqref{eqx} are located at $z_x=0$, $z_x=1$, and $z_x=z_{ax}$ with $\lvert z_{ax}\rvert>1$. Thus the matching point lies safely inside the regular interval between $z_x=0$ and $z_x=1$, making it a natural and convenient choice.

\subsubsection{Radial Heun problem and matching}
As shown in Fig.~\ref{fig:coorpatch}, the entire scattering problem cannot conveniently be covered by a single coordinate patch. We thus choose the $y$ patch and the $\bar{r}$ patch to cover it. The radial equations in these two patches are given by Eqs.~\eqref{eqy} and~\eqref{eqr}, respectively. We then recast both equations into Heun form, and impose the boundary conditions together with the connection condition to derive the relation that $\omega$ and $\lambda$ must satisfy.

We first focus on the radial equation~\eqref{eqy} in the bridge coordinate $y=-r_{\text{o}}/r$, which can be recast to a Heun equation
\begin{eqnarray}\label{heuny}
	\frac{\mathrm{d}^2Y}{\mathrm{d}z_y^2}
	+\left(\frac{\gamma_y}{z_y}+\frac{\delta_y}{z_y-1}
	+\frac{\epsilon_y}{z_y-z_{ay}}\right)
	\frac{\mathrm{d}Y}{\mathrm{d}z_y}+\frac{\alpha_y\beta_y z_y-q_y}{z_y(z_y-1)(z_y-z_{ay})}Y=0\,,
\end{eqnarray}
where
\begin{eqnarray}\label{heunyp}
	z_y&=&\frac{y_{\text{i}}-y_{-}}{y_{\text{i}}-y_{\text{o}}}
	\frac{y-y_{\text{o}}}{y-y_{-}}\,,\nonumber\\
	Y&=&(z_y-z_{0y})^{-1}z_y^{\frac{1-\gamma_y}{2}}
	(z_y-1)^{\frac{1-\delta_y}{2}}(z_y-z_{ay})^{\frac{1-\epsilon_y}{2}}\Phi^{(y)}(y)\,,\nonumber\\
	z_{ay}&=&\frac{y_{\text{i}}-y_{-}}{y_{\text{i}}-y_{\text{o}}}
	\frac{y_+-y_{\text{o}}}{y_+-y_{-}}\,,\nonumber\\ 
    z_{0y}&=&\frac{y_{\text{i}}-y_{-}}{y_{\text{i}}-y_{\text{o}}}\frac{y_{\text{o}}}{y_{-}}\,,\nonumber\\
	\gamma_y&=&1-\mathrm{i}\frac{\omega-m\varpi_{\text{o}}}{\kappa_{\text{o}}}\,,\nonumber\\ 
    \delta_y&=&1+\mathrm{i}\frac{\omega-m\varpi_{\text{i}}}{\kappa_{\text{i}}}\,,\nonumber\\
    \epsilon_y&=&1+\mathrm{i}\frac{\omega-m\varpi_+}{\kappa_+}\,,\nonumber\\
	\alpha_y&=&\mathrm{i}\frac{\omega-m\varpi_-}{2\kappa_-}
	+1+\frac{\gamma_y-1}{2}+\frac{\delta_y-1}{2}
	+\frac{\epsilon_y-1}{2}\,,\nonumber\\
	\beta_y&=&-\mathrm{i}\frac{\omega-m\varpi_-}{2\kappa_-}
	+1+\frac{\gamma_y-1}{2}+\frac{\delta_y-1}{2}
	+\frac{\epsilon_y-1}{2}\,,\nonumber \\
	q_y&=&\frac{z_{ay}}{z_{0y}}+\frac{\gamma_y\epsilon_y-1}{2}+z_{ay}\frac{\gamma_y\delta_y-1}{2}+\frac{r_+}{4}\left(1+B^2r_{\text{o}}^2\right)\Bigg[\left(
	\frac{2r_{\text{o}}}{\widehat{\rho}_0^2(y_{\text{o}})}
	-\frac{1}{r_{\text{o}}-r_{\text{i}}}
	-\frac{1}{r_{\text{o}}-r_+}
	\right)(\gamma_y-1)^2\nonumber\\
	&+&\frac{\mathrm{i}m\varpi'_{\text{o}}(\gamma_y-1)}{\kappa_{\text{o}}}
	+\frac{1}{\kappa_{\text{o}}\widehat{\rho}_0^2(y_{\text{o}})}
	\left(
	\lambda-2B^2(\mu r_{\text{o}}-a^2)
	\right)
	\Bigg]\,.
\end{eqnarray}
Here
\begin{eqnarray}
	\varpi=\frac{a}{C\rho_0^2}\,,
\end{eqnarray}
and $$\varpi'_{\text{o}}\equiv\left.\frac{\mathrm{d}\varpi}{\mathrm{d}r}\right|_{r=r_{\text{o}}}=-2\frac{r_{\text{o}}\varpi_{\text{o}}}{\widehat{\rho}_0^2(y_{\text{o}})}\, .$$ The Fuchs relation is also satisfied:
\begin{eqnarray}
	\alpha_y+\beta_y-\gamma_y-\delta_y-\epsilon_y+1=0\,.\nonumber
\end{eqnarray}
The physical ingoing boundary condition at the outer horizon, $y=y_{\text{o}}$, is imposed by choosing the local Heun branch regular at $z_y=0$:
\begin{eqnarray}\label{wavefuny}
	\Phi^{(y)}_{\text{in}}(y)&=& \mathcal{N}_{\text{in}}
	(z_y-z_{0y})z_{y}^{\frac{\gamma_y-1}{2}}(z_y-1)^{\frac{\delta_y-1}{2}}
	(z_y-z_{ay})^{\frac{\epsilon_y-1}{2}} H\ell(z_{ay},q_y;\alpha_y,\beta_y,
	\gamma_y,\delta_y;z_y)\,,
\end{eqnarray}
where
\begin{eqnarray}
	\mathcal{N}_{\text{in}}&=&\left[\frac{1}{y_{\text{o}}}(-z_{0y})\,
	\mathrm{e}^{-\mathrm{i}\pi\frac{\delta_y-1}{2}}
	(-z_{ay})^{\frac{\epsilon_y-1}{2}}
	\zeta_{\text{o}}^{\frac{\gamma_y-1}{2}}\right]^{-1}\,,\\
	\zeta_{\text{o}}&=&\lim_{r\to r_{\text{o}}}z_y\,\mathrm{e}^{-2\kappa_{\text{o}}r_*}\,.\nonumber
\end{eqnarray}
With this normalization, $\Phi^{(r)}_{\text{in}}=\Phi^{(y)}_{\text{in}}/y$ approaches exactly the unit-amplitude plane wave $\mathrm{e}^{-\mathrm{i}(\omega-m\varpi_{\text{o}})r_*}$ as $r_*\to-\infty$, and $z_y$ should be understood as a function of $y$ as shown in Eq.~\eqref{heunyp}.

We then focus on the radial equation in the $\bar{r}$ patch, where the radial equation is the same as that in the $r$ coordinate~\eqref{eqr}, and can also be recast as a Heun equation,
\begin{eqnarray}\label{heunr}
	\frac{\mathrm{d}^2\bar{R}}{\mathrm{d}z_{\bar{r}}^2}
	+\left(\frac{\gamma_{\bar{r}}}{z_{\bar{r}}}
	+\frac{\delta_{\bar{r}}}{z_{\bar{r}}-1}
	+\frac{\epsilon_{\bar{r}}}{z_{\bar{r}}-z_{a\bar{r}}}\right)
	\frac{\mathrm{d}\bar{R}}{\mathrm{d}z_{\bar{r}}}+\frac{\alpha_{\bar{r}}\beta_{\bar{r}} z_{\bar{r}}-q_{\bar{r}}}
	{z_{\bar{r}}(z_{\bar{r}}-1)(z_{\bar{r}}-z_{a\bar{r}})}\bar{R}&=&0\,,
\end{eqnarray}
where
\begin{eqnarray}\label{heunrp}
	z_{\bar{r}}&=&\frac{r_{\text{o}}-r_-}{r_{\text{o}}-r_{\text{i}}}
	\frac{\bar{r}-r_{\text{i}}}{\bar{r}-r_-}\,,\nonumber\\
	\bar{R}&=&(z_{\bar{r}}-z_{0\bar{r}})^{-1}
	z_{\bar{r}}^{\frac{1-\gamma_{\bar{r}}}{2}}
	(z_{\bar{r}}-1)^{\frac{1-\delta_{\bar{r}}}{2}}
	(z_{\bar{r}}-z_{a\bar{r}})^{\frac{1-\epsilon_{\bar{r}}}{2}}
	\bar{\Phi}^{(\bar{r})}\,,\nonumber\\
	z_{a\bar{r}}&=&\frac{r_{\text{o}}-r_-}{r_{\text{o}}-r_{\text{i}}}
	\frac{r_+-r_{\text{i}}}{r_+-r_-}\,,\nonumber\\
	z_{0\bar{r}}&=&\frac{r_{\text{o}}-r_-}{r_{\text{o}}-r_{\text{i}}}
	\frac{r_{\text{i}}}{r_-}\,,\nonumber\\
	\gamma_{\bar{r}}&=&1+\mathrm{i}\frac{\omega-m\varpi_{\text{i}}}{\kappa_{\text{i}}}\,,\nonumber\\
	\delta_{\bar{r}}&=&1-\mathrm{i}\frac{\omega-m\varpi_{\text{o}}}{\kappa_{\text{o}}}\,,\nonumber\\
    \epsilon_{\bar{r}}&=&1+\mathrm{i}\frac{\omega-m\varpi_+}{\kappa_+}\,,\nonumber\\
	\alpha_{\bar{r}}&=&\mathrm{i}\frac{\omega-m\varpi_-}{2\kappa_-}
	+1+\frac{\gamma_{\bar{r}}-1}{2}+\frac{\delta_{\bar{r}}-1}{2}
	+\frac{\epsilon_{\bar{r}}-1}{2}\,,\nonumber\\
	\beta_{\bar{r}}&=&-\mathrm{i}\frac{\omega-m\varpi_-}{2\kappa_-}
	+1+\frac{\gamma_{\bar{r}}-1}{2}+\frac{\delta_{\bar{r}}-1}{2}
	+\frac{\epsilon_{\bar{r}}-1}{2}\,,\nonumber \\
	q_{\bar{r}}&=&\frac{z_{a\bar{r}}}{z_{0\bar{r}}}
	+\frac{\gamma_{\bar{r}}\epsilon_{\bar{r}}-1}{2}+z_{a\bar{r}}\frac{\gamma_{\bar{r}}\delta_{\bar{r}}-1}{2}+\frac{1}{4}r_+(1+B^2r_{\text{i}}^2)\Bigg[
	\left(
	\frac{2r_{\text{i}}}{\rho_0^2(r_{\text{i}})}-\frac{1}{r_{\text{i}}-r_{\text{o}}}-\frac{1}{r_{\text{i}}-r_+}
	\right)(\gamma_{\bar{r}}-1)^2\nonumber\\
	&-&\frac{\mathrm{i}m\varpi'_{\text{i}}(\gamma_{\bar{r}}-1)}{\kappa_{\text{i}}}+\frac{1}{\kappa_{\text{i}}\,\rho_0^2(r_{\text{i}})}
	\left(
	\lambda+\frac{2(\mu r_{\text{i}}-a^2)}{r_{\text{i}}^2}
	\right)
	\Bigg]\,.
\end{eqnarray}
Here $$\varpi'_{\text{i}}\equiv\left.\frac{\mathrm{d}\varpi}{\mathrm{d}r}\right|_{r=r_{\text{i}}}=-2\frac{r_{\text{i}}\varpi_{\text{i}}}{\rho_0^2(r_{\text{i}})}\, .$$It immediately follows that the Fuchs relation holds:
\begin{eqnarray}
	\alpha_{\bar{r}}+\beta_{\bar{r}}-\gamma_{\bar{r}}-\delta_{\bar{r}}-\epsilon_{\bar{r}}+1=0\,.\nonumber
\end{eqnarray}

Two independent local solutions at the inner horizon of the second universe, $z_{\bar{r}}=0$ or equivalently $\bar{r}=\bar{r}_{\mathrm{i}}$, are
\begin{equation}\label{wavefunr}
	\begin{aligned}
		\bar{\Phi}_{\text{up}}^{(\bar{r})}
		=      &
		\mathcal{N}_{\text{up}}
		(z_{\bar{r}}-z_{0\bar{r}})
		z_{\bar{r}}^{\frac{\gamma_{\bar{r}}-1}{2}}
		(z_{\bar{r}}-1)^{\frac{\delta_{\bar{r}}-1}{2}}
		(z_{\bar{r}}-z_{a\bar{r}})^{\frac{\epsilon_{\bar{r}}-1}{2}}
		H\ell\left(
		z_{a\bar{r}},q_{\bar{r}};
		\alpha_{\bar{r}},\beta_{\bar{r}},\gamma_{\bar{r}},
		\delta_{\bar{r}};z_{\bar{r}}
		\right),\nonumber\\
		\bar{\Phi}_{\text{down}}^{(\bar{r})}
		=      &
		\mathcal{N}_{\text{down}}
		(z_{\bar{r}}-z_{0\bar{r}})
		z_{\bar{r}}^{\frac{1-\gamma_{\bar{r}}}{2}}
		(z_{\bar{r}}-1)^{\frac{\delta_{\bar{r}}-1}{2}}
		(z_{\bar{r}}-z_{a\bar{r}})^{\frac{\epsilon_{\bar{r}}-1}{2}}
		H\ell\left(
		z_{a\bar{r}},q_{1\bar{r}};
		\alpha_{\bar{r}}+1-\gamma_{\bar{r}},
		\beta_{\bar{r}}+1-\gamma_{\bar{r}},
		2-\gamma_{\bar{r}},\delta_{\bar{r}};z_{\bar{r}}
		\right),
	\end{aligned}
\end{equation}
with
\begin{eqnarray}
	q_{1\bar{r}}&=&q_{\bar{r}}-(\gamma_{\bar{r}}-1)\left(\epsilon_{\bar{r}}+z_{a\bar{r}}\delta_{\bar{r}}\right)\,,\nonumber\\
	\mathcal{N}_{\text{up}}&=&\left[
	(-z_{0\bar{r}})\,\mathrm{e}^{+\mathrm{i}\pi\frac{\delta_{\bar{r}}-1}{2}}
	(-z_{a\bar{r}})^{\frac{\epsilon_{\bar{r}}-1}{2}}
	\zeta_{\text{i}}^{\frac{\gamma_{\bar{r}}-1}{2}}\right]^{-1}\,,\nonumber\\
	\mathcal{N}_{\text{down}}&=&\left[
	(-z_{0\bar{r}})\,\mathrm{e}^{+\mathrm{i}\pi\frac{\delta_{\bar{r}}-1}{2}}
	(-z_{a\bar{r}})^{\frac{\epsilon_{\bar{r}}-1}{2}}
	\zeta_{\text{i}}^{\frac{1-\gamma_{\bar{r}}}{2}}\right]^{-1}\,,\nonumber\\
	\zeta_{\text{i}}&=&\lim_{\bar r\to \bar{r}_{\text{i}}^-}z_{\bar r}\,\mathrm{e}^{-2\kappa_{\text{i}}\bar r_*}\,.\nonumber
\end{eqnarray}
Here $z_{\bar r}=z_{\bar r}(\bar r)$ is defined in Eq.~\eqref{heunrp}. The solutions $\bar{\Phi}_{\mathrm{up}}^{(\bar r)}$ and $\bar{\Phi}_{\mathrm{down}}^{(\bar r)}$ satisfy the outgoing and ingoing boundary conditions at $r_*\to+\infty$, respectively, and are normalized to have unit amplitude: $\bar{\Phi}_{\mathrm{up}}^{(\bar r)}\to\mathrm{e}^{+\mathrm{i}(\omega-m\varpi_{\mathrm{i}})r_*}$ and $\bar{\Phi}_{\mathrm{down}}^{(\bar r)}\to\mathrm{e}^{-\mathrm{i}(\omega-m\varpi_{\mathrm{i}})r_*}$.

Now, we have two local solutions, \eqref{wavefuny} and \eqref{wavefunr}, each of which satisfies the boundary conditions only partially, in the $y$ patch and the $\bar{r}$ patch, respectively. We then match them at a common overlap point in the second universe, described by $y_\text{m}\in(0,+\infty)$ in the $y$ patch and by $\bar{r}_\text{m}=-r_{\text{o}}/y_\text{m}\in(-\infty,0)$ in the $\bar{r}$ patch, as schematically labeled in Fig.~\ref{fig:coorpatch}. In principle, the matching point can be chosen arbitrarily within the overlap region, provided that the wave functions are analytically continued consistently. In practice, however, after the matching point is mapped to the complex $z$-plane, the continuation path from the physical boundary to the matching point in a given coordinate patch may cross the conventional branch cut of the Heun function. Such a crossing introduces a spurious jump in the wave function and should therefore be avoided by a careful choice of the matching point. Branch switches associated with the elementary prefactors multiplying the Heun function are ignored here, as they amount only to an overall multiplicative factor, which does not affect the search of the QNMs.

\begin{figure}[htbp]
	\centering
	\resizebox{0.5\columnwidth}{!}{%
		\includegraphics{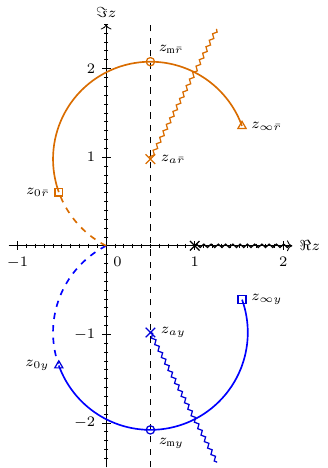}}
	\caption{\justifying
		Trajectories of the images of the matching point, $z_{\text{m}y}$ and $z_{\text{m}\bar{r}}$, in the complex $z$-plane. They are shown as blue and orange circular arcs centered exactly at $z_{ay}$ and $z_{a\bar{r}}$, respectively. The zigzag lines denote branch cuts, while the dashed lines indicate segments with no corresponding points in the other patch. Crosses mark the poles; hollow squares mark $z_{\infty y}$ and $z_{0\bar{r}}$; hollow triangles mark $z_{0y}$ and $z_{\infty\bar{r}}$; and hollow circles mark $z_{\text{m}y}$ and $z_{\text{m}\bar{r}}$ for the choice $y_{\text{m}}=B^2 M r_{\text{o}}/I_1$. The local solutions are analytically continued from $z=0$ along the circular arcs. To avoid crossing branch cuts, the allowed range of the matching point $y_{\text{m}}$ is restricted as in Eq.~\eqref{ma}. The parameters are chosen as $M=1$, $a=0.9$, and $B=0.6$ for this figure.
	}\label{fig:ma}
\end{figure}

The branch cuts of a Heun function are taken to be the half-line starting at $z=1$ and extending along the real axis towards positive infinity, and the half-line starting at $z=z_a$ and extending to infinity along the direction $\arg z_a$. The singularities of the two Heun equations, Eq.~\eqref{eqy} and Eq.~\eqref{eqr}, are mapped to $z_y=0, z_y=1, z_y=z_{ay}$ and $z_{\bar{r}}=0, z_{\bar{r}}=1, z_{\bar{r}}=z_{a\bar{r}}$, respectively. To choose an appropriate matching point, one should inspect how $z_{\text{m}y}=z_{y}(y_\text{m})$ and $z_{\text{m}\bar{r}}=z_{\bar{r}}(\bar{r}_\text{m})$ move as $y_\text{m}$ varies, which is shown in Fig.~\ref{fig:ma}. It can be shown that $$z_{y}(y)+z_{\bar{r}}(-r_{\text{o}}/y)=1\,,$$ so the pair $z_{\text{m}y}, z_{\text{m}\bar{r}}$ and the pair $z_{ay}, z_{a\bar{r}}$ are both symmetric with respect to $z=1/2$. Additionally, it can be proved that $\Re z_{ay}=\Re z_{a\bar{r}}=1/2$, and the trajectories of $z_{\text{m}y}$ and $z_{\text{m}\bar{r}}$ are circles whose centers are exactly $z_{ay}$ and $z_{a\bar{r}}$, respectively. As $y_\text{m}$ moves from $0$ to $+\infty$, the $z_{\text{m}y}$ appears in the lower half of the complex $z$-plane and runs from $z_{0y}$ to $z_{\infty y}$, while the $z_{\text{m}\bar{r}}$ appears in the upper half of the complex $z$-plane and runs from $z_{\infty \bar{r}}$ to $z_{0\bar{r}}$. The dashed lines label the segments $-1<y_\text{m}<0$ and $0<\bar{r}_\text{m}<r_{\text{i}}$, respectively, where there are no corresponding points in the other patch. The zigzag lines label branch cuts. The local solutions, \eqref{wavefuny} and \eqref{wavefunr}, are analytically continued from their respective physical boundaries along real $y$ and $\bar{r}$. Equivalently, in the complex-$z$ plane, the continuation starts from $z=0$ and follows the corresponding circular arcs. To avoid crossing the branch cuts, the matching point is restricted to
\begin{eqnarray}\label{ma}
	B^2r_\text{o}r_\text{i}<y_\text{m}<B^2r_\text{o}^2\,.
\end{eqnarray}
We can parametrize the matching point as $$y_\text{m}=B^2r_\text{o}\left[\xi r_\text{i}+(1-\xi)r_\text{o}\right]$$ with $\xi\in(0,1)$. In practice, we test matching points corresponding to several $\xi$, as well as the special choice $y_\text{m}=B^2Mr_\text{o}/I_1$. The latter yields $\Re z_{\text{m}y}=\Re z_{\text{m}\bar{r}}=1/2$. These choices lead to different computational costs but yield the same QNM frequencies.

QNMs are determined by requiring that the incident coefficient $A_{\text{in}}$ at the matching point vanish
\begin{eqnarray}\label{raCon}
	A_{\text{in}}=\left.\frac{W\left[\bar{\Phi}_{\text{up}}^{(\bar{r})},\Phi_{\text{in}}^{(y)}/y\right]}{W\left[\bar{\Phi}_{\text{up}}^{(\bar{r})},\bar{\Phi}_{\text{down}}^{(\bar{r})}\right]}\right|_{
	\begin{subarray}
		\bar{r}  = \bar{r}_{\mathrm{m}}, \\
		y        = y_{\mathrm{m}}
	\end{subarray}}=0\,,
\end{eqnarray}
in which the derivative is taken consistently with respect to $\bar r$. This Wronskian ratio is independent of the evaluation point, as long as all solutions are analytically continued consistently. This condition selects a set of $(\omega,\lambda)$ such that the local solutions can be smoothly extended into another coordinate patch without developing unphysical branches, thereby forming a global radial solution that satisfies the boundary conditions at both ends of the scattering domain. Together with the condition from the angular part, Eq.~\eqref{anCon}, it then determines a discrete set of quasinormal frequencies and the associated separation constants.

\section{Numerical Results}\label{sec-spectra}
We set $M=1$ and focus on the scalar $(\ell,m)=(2,2)$ sector. The numerical survey varies both $BM$ at fixed rotation and $a/M$ at fixed electromagnetic field, subject to subextremality and $A>0$. Even in the weak-field regime the spectrum does not approach the Kerr QNM spectrum. The reason is global rather than perturbative. The scattering domain continues across the wormhole-like bridge to the neighboring inner horizon instead of ending at Kerr null infinity.

\begin{figure}[htbp]
    \centering
    \includegraphics[width=0.49\linewidth]{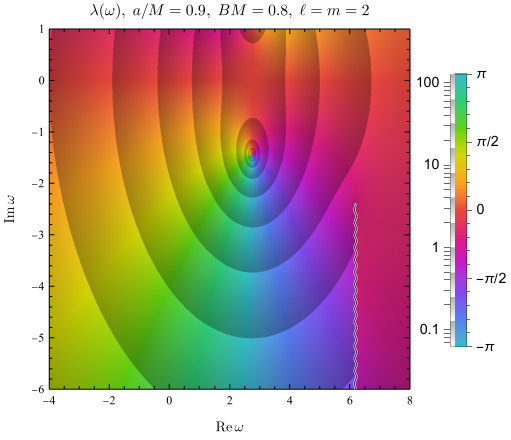}
    \includegraphics[width=0.49\linewidth]{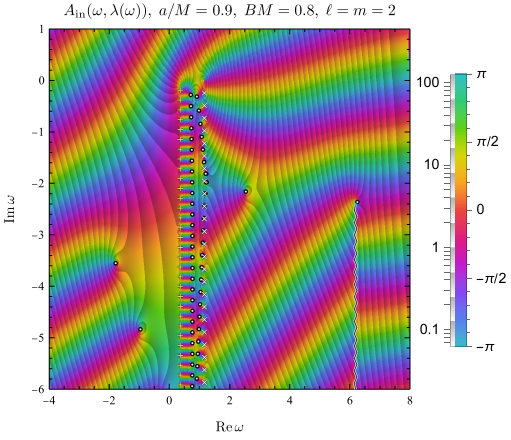}
    \caption{\justifying The angular eigenvalue $\lambda(\omega)$ (left) and the incident amplitude $A_{\mathrm{in}}(\omega,\lambda(\omega))$ (right) in the complex $\omega$ plane for $a/M=0.9$, $BM=0.8$, and $\ell=m=2$. The colors encode the complex phase, while the repeated light--dark bands indicate the magnitude. Black circles mark the zeros of $A_{\mathrm{in}}$ and hence the QNM frequencies. Plus signs and crosses mark the poles of $A_{\mathrm{in}}$. They correspond to the Matsubara modes of the outer horizon of the first KBR region and the inner horizon of the neighboring region, respectively: $\omega_n^{(\mathrm{o})}=m\varpi_{\mathrm{o}}-\mathrm{i}(n+1)\kappa_{\mathrm{o}}$ and $\omega_n^{(\mathrm{i})}=m\bar{\varpi}_{\mathrm{i}}-\mathrm{i}(n+1)|\bar{\kappa}_{\mathrm{i}}|$, with $n=0,1,2,\ldots$. Each family forms an exactly vertical ladder; the slight apparent curvature is an optical illusion caused by the surrounding QNMs. The zigzag line denotes a chosen branch cut emanating from the exceptional point of $\lambda(\omega)$, where the same-parity $\ell=2$ and $\ell=4$ branches coalesce; this branch structure is inherited by $A_{\mathrm{in}}(\omega,\lambda(\omega))$. No additional branch cut emanating from $\omega=0$ appears in $A_{\mathrm{in}}(\omega,\lambda(\omega))$, as expected since neither scattering boundary is asymptotically flat null infinity.
}
    \label{fig:Ain}
\end{figure}

To initialize the spectral calculation, we first identify the angular branch corresponding to the desired $\ell$. For each chosen $a/M$, the appropriate eigenvalue $\lambda$ is selected near $\omega=0$ in the weak $BM$ regime, where the $\ell$ branches can be identified unambiguously, and is then continuously tracked to the target $BM$. Then the same branch $\lambda(\omega)$ is continued over the relevant region of the complex-$\omega$ plane. We then perform a two-dimensional scan of $A_{\mathrm{in}}(\omega,\lambda(\omega))$, whose zeros are QNMs. These zeros are then subsequently refined numerically. Once identified, these modes serve as seeds for continuation in $a/M$ or $BM$, allowing the corresponding QNM branches to be tracked through parameter space. A representative example for $a/M=0.9$, $BM=0.8$, and $\ell=m=2$ is shown in Fig.~\ref{fig:Ain}.

\begin{figure*}[htbp]
\centering
\begin{subfigure}{0.45\textwidth}
\centering
\includegraphics[width=\linewidth]{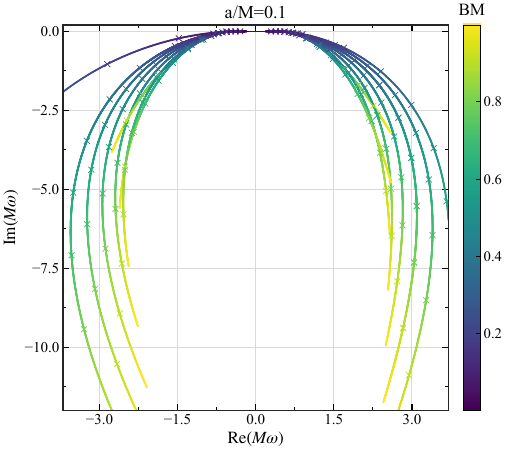}
\end{subfigure}\hfill
\begin{subfigure}{0.45\textwidth}
\centering
\includegraphics[width=\linewidth]{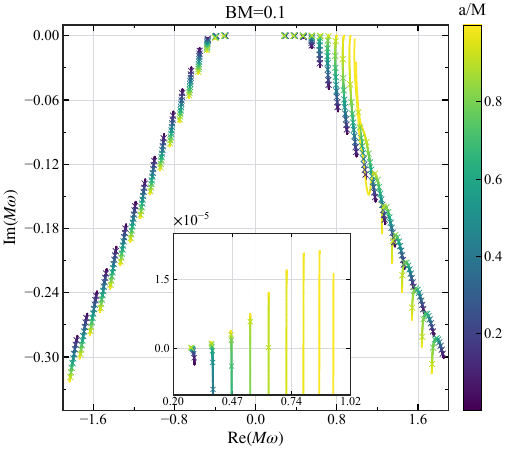}
\end{subfigure}\\

\begin{subfigure}{0.45\textwidth}
\centering
\includegraphics[width=\linewidth]{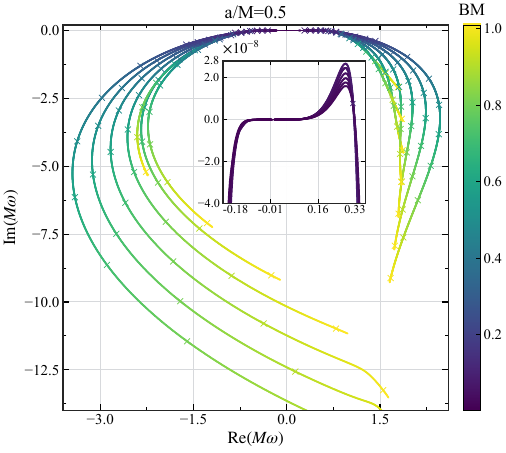}
\end{subfigure}\hfill
\begin{subfigure}{0.45\textwidth}
\centering
\includegraphics[width=\linewidth]{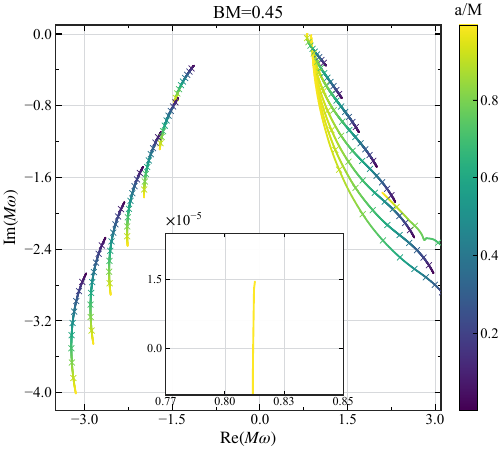}
\end{subfigure}\\

\begin{subfigure}{0.45\textwidth}
\centering
\includegraphics[width=\linewidth]{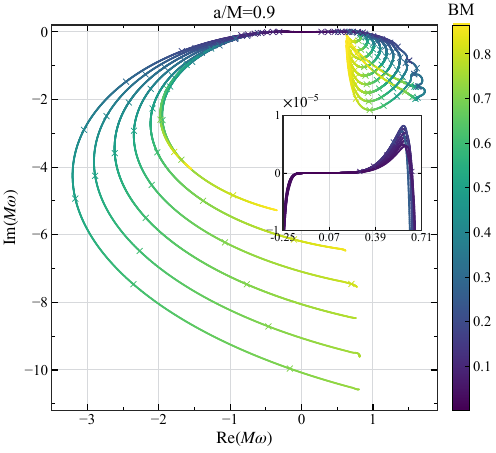}
\end{subfigure}\hfill
\begin{subfigure}{0.45\textwidth}
\centering
\includegraphics[width=\linewidth]{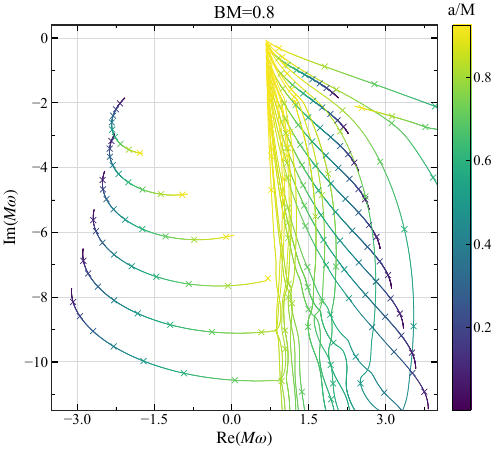}
\end{subfigure}
\caption{\justifying Full scalar $(\ell,m)=(2,2)$ QNM spectra for representative parameter scans. The left column varies $BM$ at fixed $a/M$, while the right column varies $a/M$ at fixed $BM$. The panels display the parameter dependence used to identify the unstable regime at sufficiently rapid rotation and sufficiently weak electromagnetic field.}
\label{fig:other-qnms}
\end{figure*}

Figure~\ref{fig:other-qnms} shows the QNM trajectories for representative scans. Several branches cross into the upper half of the complex-frequency plane. The common trend is that unstable modes are favored by sufficiently large $a/M$ and sufficiently small $BM$. Proximity to extremality by itself is therefore not the appropriate characterization of the unstable regime.

The unstable nonaxisymmetric branches cross the real axis at marginal modes. There, $\omega$ and the angular eigenvalue are real, so the radial equation has real coefficients and the $r_*$-Wronskian of an eigenfunction and its complex conjugate is conserved. Evaluating it at the outer horizon of the first universe and the inner horizon of the neighboring universe with the QNM boundary conditions gives
\begin{equation}
(\omega-m\varpi_{\mathrm o})|A_{\mathrm o}|^2+(\omega-m\bar\varpi_{\mathrm i})|A_{\mathrm i}|^2=0\,.
\end{equation}
This is the conserved radial-wave current, while multiplication by $\omega$ gives the corresponding $\partial_t$ Killing-energy balance. Since $m\varpi_{\mathrm o}<m\bar\varpi_{\mathrm i}$, any nontrivial marginal mode satisfies
\begin{equation}
m\varpi_{\mathrm o}<\omega<m\bar\varpi_{\mathrm i}\,.
\end{equation}
With the adopted future-horizon and radial-flux orientations, the outer-horizon term therefore represents absorption and the inner-horizon term superradiant extraction. The inequality follows directly from the Wronskian balance rather than providing an independent numerical signature.

Superradiant extraction alone does not produce an instability. It requires feedback that continually returns incident flux to the amplifying boundary. The double-barrier potential identified in Sec.~\ref{sec-potential} supplies such a mechanism. A trapped wave can leak through the right barrier toward the neighboring inner horizon, couple to the negative-Killing-energy branch, and return an amplified reflected component to the cavity. The unstable QNM branches emerge continuously from the marginal real modes as the parameters are varied. This provides strong evidence that the nonaxisymmetric instability is a black-hole-bomb mechanism~\cite{Press:1972zz, Cardoso:2004nk} generated by superradiant amplification together with trapping.

The same cavity exists more broadly than the unstable regime. The spectra contain many weakly damped branches whose trajectories lie close to and approximately parallel to the real axis. Such long-lived resonances are characteristic of a cavity bounded by two partially transmitting barriers. They suggest repeated reflections across the gluing region and hence echo-like late-time responses. The echo interpretation and the instability should therefore be distinguished. The former follows from long-lived trapping, while the latter additionally requires net amplification over a feedback cycle.

The axisymmetric and nonaxisymmetric growing modes also have different physical origins. The $m=0$ modes are tied to the chronology-violating region and arise within a scattering domain that is not globally hyperbolic. The $m=2$ branches instead probe rotational dynamics and exhibit a clear marginal superradiant balance. Neither result alone proves nonlinear instability of the full Einstein-Maxwell solution. Gravitational and electromagnetic perturbations, and ultimately nonlinear evolution where an initial-value formulation is available, are required to address that question.

\section{Discussion and Conclusions}\label{sec-conclusion}

We have constructed the natural extension of the Kerr-Bertotti-Robinson spacetime across the Boyer-Lindquist-type surface $r=\infty$. The surface $r=+\infty$ of one KBR region is smoothly connected to $\bar r=-\infty$ of a neighboring region, and repeated continuation produces an infinite chain of regions joined by wormhole-like bridges. This extension is not restricted to the generic subextremal case: the causal character of the gluing hypersurface changes across parameter space, being timelike for $A>0$, null for $A=0$, and spacelike for $A<0$, with the extremal and static sectors admitting the corresponding limiting constructions. The surface $r=\infty$ should therefore be regarded as an internal continuation surface rather than an asymptotic boundary.

This global extension changes the physical interpretation of the KBR geometry. In particular, the neighboring negative-radius region, including its chronology-violating sector and ring singularity, becomes directly connected to the original exterior without an intervening horizon. The exposed singularity challenges weak cosmic censorship conjecture and motivates us to investigate the stability of the extended geometry through its QNM spectrum.

The scalar spectrum provides a complementary probe of this extended structure. In the axisymmetric sector, purely imaginary growing QNMs exist for every $\ell$, with the associated trapping region tied to the chronology-violating sector. Since the scattering region is not globally hyperbolic, these modes do not establish conventional dynamical instability arising from generic Cauchy data. A distinct family of unstable modes appears in the nonaxisymmetric $(\ell,m)=(2,2)$ sector for sufficiently large $a/M$ and sufficiently small $BM$. At the marginal real modes, the conserved radial Wronskian gives an exact balance between outer-horizon absorption and superradiant extraction from the neighboring inner horizon. Together with the double-barrier cavity and the continuous emergence of unstable branches from these marginal modes, this provides strong evidence for a black-hole-bomb mechanism driven by superradiant amplification and trapping. 

The absence of a conventional null infinity precludes the extended KBR spacetime from being regarded as an isolated black hole in the usual global sense. Its outer Killing horizon nevertheless remains well defined and admits a natural quasi-local interpretation. The extension therefore provides a setting in which familiar horizon phenomena coexist with a global structure qualitatively different from that of Kerr. This global structure also leaves a characteristic imprint on wave propagation: the double-barrier cavity spanning the gluing region supports families of weakly damped modes and may give rise to echo-like responses.

The present analysis is restricted to a test massless scalar field. An important next step is therefore to study the coupled gravitational and electromagnetic perturbations and determine whether analogous growing and long-lived modes persist in the physical perturbations of the Einstein-Maxwell system. The global structure associated with the $\Omega=0$ locus, and in particular the appropriate description of null infinity, also remains to be understood.

Following the appearance of our short report~\cite{Zhou:2026tkm}, and while the present detailed study was being completed, the broader Kerr-Newman-Bertotti-Robinson family was identified~\cite{Ovcharenko:2026tos}. The extension method developed here generalizes naturally to the generic KNBR family, as also demonstrated in the parallel work~\cite{Ovcharenko:2026ooh}. Special parameter sectors, however, may require separate treatment. These sectors, together with possible alternative extensions and the global structure of null infinity in the KNBR spacetime, are currently under investigation.

\section*{Acknowledgement}
We thank Hryhorii Ovcharenko for bringing his parallel work to our attention after the completion of our earlier short report and for the helpful discussions. This work is supported by the National Key R\&D Program of China Grant No. 2022YFC2204603, and by the National Natural Science Foundation of China with grants No. 12475063, No. 12247103, No. 12505067, No. 12588101, and No. 12535002.

\bibliography{mainRef.bib}
\bibliographystyle{apsrev4-1}

\end{document}